\documentclass[sigconf,screen]{acmart} % authorversion

\newcommand{\nparindent}{\hspace{10pt}}


\AtBeginDocument{%
  \providecommand\BibTeX{{%
    \normalfont B\kern-0.5em{\scshape i\kern-0.25em b}\kern-0.8em\TeX}}}

\setcopyright{acmlicensed}
\acmPrice{15.00}
\acmDOI{10.1145/3368089.3409710}
\acmYear{2020}
\copyrightyear{2020}
\acmSubmissionID{fse20main-p315-p}
\acmISBN{978-1-4503-7043-1/20/11}
\acmConference[ESEC/FSE '20]{Proceedings of the 28th ACM Joint European Software Engineering Conference and Symposium on the Foundations of Software Engineering}{November 8--13, 2020}{Virtual Event, USA}
\acmBooktitle{Proceedings of the 28th ACM Joint European Software Engineering Conference and Symposium on the Foundations of Software Engineering (ESEC/FSE '20), November 8--13, 2020, Virtual Event, USA}

\usepackage{wrapfig}
\usepackage{tikz}
\usetikzlibrary{decorations.text}
\usepackage{float}
\usepackage{microtype}

\newcommand{\approachname}[0]{Non-Optimizing Reference Engine Construction}
\newcommand{\approachnameshort}[0]{NoREC}

\newcommand{\bugsymbol}[0]{\includegraphics[width=0.3cm]{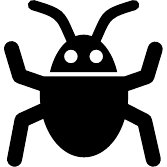}}
\newcommand{\oksymbol}[0]{\includegraphics[width=0.3cm]{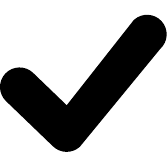}}

\newcommand{\citequerygenerators}[0]{\cite{sqlsmith,qgen,Bati2007,apollo,Bruno2006, qrelx, Mishra2008}}
\newcommand{\citedbgenerators}[0]{\cite{qagen,khalek2008,gray1994,houkjaer2006,nicolas2005,neufeld1993}}
\newcommand{\ie}{\emph{i.e.}}
\newcommand{\eg}{\emph{e.g.}}
\newcommand{\cf}{\emph{cf.}}

\usepackage{xcolor,listings}

\usepackage{graphicx}
\usepackage{balance}  % for  \balance command ON LAST PAGE  (only there!)
\usepackage{booktabs}
\newcommand{\tablesize}{\small{}}
\newcommand{\sqlite}{SQLite}
\newcommand{\postgres}{PostgreSQL}
\newcommand{\reducespace}{\vspace{-1.2em}}
\usepackage{array}
\usepackage[originalparameters]{ragged2e}  % The option prevents hyphentation rather than making it rare 
\usepackage{tikz}
\newcommand*\circled[1]{\tikz[baseline=(char.base)]{
            \node[shape=circle,draw,inner sep=1pt] (char) {#1};}}
\usepackage{float}

\begin{document}

\title[Detecting Optimization Bugs in Database Engines via \approachnameshort{}]{Detecting Optimization Bugs in Database Engines via \approachname{}}

\author{Manuel Rigger}
\email{manuel.rigger@inf.ethz.ch}
\affiliation{%
  \institution{Department of Computer Science, ETH Zurich}
  \city{Zurich}
  \country{Switzerland}
}

\author{Zhendong Su}
\email{zhendong.su@inf.ethz.ch}
\affiliation{%
  \institution{Department of Computer Science, ETH Zurich}
  \city{Zurich}
  \country{Switzerland}
}

\begin{abstract}
Database Management Systems (DBMS) are used ubiquitously.
To efficiently access data, they apply sophisticated optimizations.
Incorrect optimizations can result in \emph{logic bugs}, which cause a query to compute an incorrect result set.
We propose \emph{\approachname{}} (\approachnameshort{}), a fully-automatic approach to detect optimization bugs in DBMS.
Conceptually, this approach aims to evaluate a query by an optimizing and a non-optimizing version of a DBMS, to then detect differences in their returned result set, which would indicate a bug in the DBMS.
Obtaining a non-optimizing version of a DBMS is challenging, because DBMS typically provide limited control over optimizations.
Our core insight is that a given, potentially randomly-generated \emph{optimized query} can be rewritten to one that the DBMS cannot optimize.
Evaluating this unoptimized query effectively corresponds to a non-optimizing reference engine executing the original query.
We evaluated \approachnameshort{} in an extensive testing campaign on four widely-used DBMS, namely PostgreSQL, MariaDB, SQLite, and CockroachDB.
We found \nrTruePositives{} previously unknown bugs in the latest versions of these systems, \fixed{} of which have been fixed by the developers.
Of these, \sumnorec{} were optimization bugs, while the remaining were error and crash bugs.
Our results suggest that \approachnameshort{} is effective, general and requires little implementation effort, which makes the technique widely applicable in practice.
\end{abstract}



%%%%%%%%%%%%% auto-generated

\newcommand{\nrBugscockroachdb}{35}
\newcommand{\nrBugsmariadb}{6}
\newcommand{\nrBugspostgres}{8}
\newcommand{\nrBugssqlite}{110}
\newcommand{\cockroachdbclosedduplicate}{1}
\newcommand{\cockroachdbclosednotabug}{0}
\newcommand{\cockroachdbfixed}{28}
\newcommand{\cockroachdbfixedindocumentation}{0}
\newcommand{\cockroachdbfixedInDocsOrCode}{28}
\newcommand{\cockroachdbopen}{0}
\newcommand{\cockroachdbverified}{7}
\newcommand{\mariadbclosedduplicate}{1}
\newcommand{\mariadbclosednotabug}{0}
\newcommand{\mariadbfixed}{1}
\newcommand{\mariadbfixedindocumentation}{0}
\newcommand{\mariadbfixedInDocsOrCode}{1}
\newcommand{\mariadbopen}{0}
\newcommand{\mariadbverified}{5}
\newcommand{\postgresclosedduplicate}{0}
\newcommand{\postgresclosednotabug}{1}
\newcommand{\postgresfixed}{5}
\newcommand{\postgresfixedindocumentation}{0}
\newcommand{\postgresfixedInDocsOrCode}{5}
\newcommand{\postgresopen}{1}
\newcommand{\postgresverified}{2}
\newcommand{\sqliteclosedduplicate}{0}
\newcommand{\sqliteclosednotabug}{6}
\newcommand{\sqlitefixed}{107}
\newcommand{\sqlitefixedindocumentation}{3}
\newcommand{\sqlitefixedInDocsOrCode}{110}
\newcommand{\sqliteopen}{0}
\newcommand{\sqliteverified}{0}
\newcommand{\closedduplicate}{2}
\newcommand{\closednotabug}{7}
\newcommand{\fixed}{141}
\newcommand{\fixedindocumentation}{3}
\newcommand{\open}{1}
\newcommand{\verified}{14}
\newcommand{\cockroachdbcrash}{4}
\newcommand{\cockroachdbdebug}{0}
\newcommand{\cockroachdberror}{24}
\newcommand{\cockroachdbnorec}{7}
\newcommand{\mariadbcrash}{1}
\newcommand{\mariadbdebug}{0}
\newcommand{\mariadberror}{0}
\newcommand{\mariadbnorec}{5}
\newcommand{\postgrescrash}{3}
\newcommand{\postgresdebug}{1}
\newcommand{\postgreserror}{4}
\newcommand{\postgresnorec}{0}
\newcommand{\sqlitecrash}{15}
\newcommand{\sqlitedebug}{26}
\newcommand{\sqliteerror}{30}
\newcommand{\sqlitenorec}{39}
\newcommand{\sumcrash}{23}
\newcommand{\sumdebug}{27}
\newcommand{\sumerror}{58}
\newcommand{\sumnorec}{51}
% overview
\newcommand{\nrTruePositives}{159}
\newcommand{\nrFalsePositives}{9}
\newcommand{\nrReported}{168}
% oracle comparison (bugs found only by NoREC)
\newcommand{\aggregate}{5}
\newcommand{\duplicates}{4}
\newcommand{\incorrectlyfetched}{13}
% oracle comparison (bugs found only by PQS)
\newcommand{\norecaggregate}{3}
\newcommand{\norecdistinct}{3}
\newcommand{\norecintersect}{1}
\newcommand{\norecleftjoin}{1}
\newcommand{\norecunknown}{18}
\newcommand{\percNoRECBugsFoundByPQS}{56.9}
\newcommand{\percNoRECBugsFoundByPQSDisregardingMistakenlyFetchedRows}{82.4}
\newcommand{\percPqsBugsFoundByNoREC}{52.7}
% tags
\newcommand{\cockroachdbCORE}{20}
\newcommand{\cockroachdbEXPERIMENTAL}{15}
\newcommand{\cockroachdbINTERNALERROR}{17}
\newcommand{\cockroachdbVECTORIZE}{11}
\newcommand{\mariadbGENERATEDCOLUMN}{1}
\newcommand{\postgresDEBUG}{1}
\newcommand{\sqliteCORE}{71}
\newcommand{\sqliteDBSTAT}{2}
\newcommand{\sqliteDEBUG}{27}
\newcommand{\sqliteFTS}{24}
\newcommand{\sqliteGENERATEDCOLUMN}{22}
\newcommand{\sqliteRTREE}{13}
% tags with false positives
\newcommand{\cockroachdbCOREwithFP}{20}
\newcommand{\cockroachdbEXPERIMENTALwithFP}{16}
\newcommand{\cockroachdbINTERNALERRORwithFP}{18}
\newcommand{\cockroachdbVECTORIZEwithFP}{12}
\newcommand{\mariadbGENERATEDCOLUMNwithFP}{1}
\newcommand{\postgresDEBUGwithFP}{1}
\newcommand{\sqliteCOREwithFP}{76}
\newcommand{\sqliteDBSTATwithFP}{3}
\newcommand{\sqliteDEBUGwithFP}{27}
\newcommand{\sqliteFTSwithFP}{24}
\newcommand{\sqliteGENERATEDCOLUMNwithFP}{22}
\newcommand{\sqliteRTREEwithFP}{13}
% misc commands
\newcommand{\maxloc}{12}
\newcommand{\avgloc}{3.82}
\newcommand{\nrtestcaseswithsingleline}{7}
\newcommand{\totalnrbugs}{159}
% sqlite severity
\newcommand{\sqliteNrCriticalBugs}{16}
\newcommand{\sqliteNrImportantBugs}{24}
\newcommand{\sqliteNrMinorBugs}{5}
\newcommand{\sqliteNrNoneBugs}{59}
\newcommand{\sqliteNrSevereBugs}{4}
\newcommand{\sqliteNrminorBugs}{2}
\newcommand{\nrBugReportsWithCREATETABLE}{116}
\newcommand{\percBugReportsWithCREATETABLE}{73.0}
\newcommand{\percOneCreateTable}{88.0}
\newcommand{\percTwoCreateTable}{12.0}
\newcommand{\mySQLNrEngineBugs}{0}
\newcommand{\mySQLNrUnsignedBugs}{0}
\newcommand{\percUNIQUE}{24.5}
\newcommand{\percPRIMARYKEY}{8.2}
\newcommand{\percINDEX}{20.1}
\newcommand{\percFOREIGNKEY}{3.1}
\newcommand{\percCOLLATE}{10.7}
\newcommand{\sqliteGeneratedColumnSegfaults}{9}

%%%%%%%%%%%%%%%%%%%%%%%%%%%%

\maketitle

\section{Introduction}
Database Management Systems (DBMS) are an important component in many systems.
To meet the growing performance demands, increasingly sophisticated optimizations for query evaluation are applied~\cite{Wu2018,Ding2018,Neumann2018,Marcus2019}.
Unsurprisingly, the \emph{query optimizer} is typically considered to be a DBMS' most complex component, posing a major correctness challenge~\cite{giakoumakis2008testing,Torsten2008}.
Implementation errors in the optimizer can result in \emph{logic bugs}, which are bugs that cause a DBMS to return an incorrect result set for a given query.
Specifically, we refer to logic bugs in the query optimizer as \emph{optimization bugs}.
Pivoted Query Synthesis (PQS) was recently proposed as a way of tackling logic bugs in DBMS~\cite{pqs}.
Its core idea is to verify the DBMS based on a single \emph{pivot row}, for which a query is generated that is expected to fetch this row.
While PQS has been effective in detecting many bugs in widely-used DBMS, a significant drawback is the high implementation effort that is required to realize this technique; specifically, the technique requires the re-implementation of the DBMS' provided operators and functions to determine whether a randomly-generated expression evaluates to \texttt{TRUE}.
Since PQS considers only a single row, it also fails to detect bugs such as when a duplicate row is mistakenly fetched or omitted.
Another successful technique for detecting logic bugs in DBMS was realized in a system called RAGS~\cite{slutz1998massive}.
It is based on differential testing~\cite{mckeeman1998differential}.
A query is generated that is sent to multiple DBMS; if the DBMS disagree on the output, at least one of the DBMS is expected to be affected by a bug.
As noted by the authors, a significant drawback of this technique is that it applies only to the common core of SQL, which is small, because DBMS differ in what operators and types they support and because even common operators have subtly different semantics between different DBMS~\cite{slutz1998massive}.

\nparindent{}In this paper, we propose \emph{\approachname{} (\approachnameshort{})}, a novel, general, and cost-effective technique for finding optimization bugs in DBMS.
The high-level idea of our approach is to compare the results of an optimizing version of a DBMS against a version of the same DBMS that does not perform any optimizations.
Obtaining such a non-optimizing version of a DBMS is challenging.
While many DBMS provide some options to control optimizations, these are limited and specific to a DBMS.
Although adding such options would be a possibility, doing so retrospectively would be error prone and impractical because of the high implementation effort and domain knowledge required.
Rather, we propose the idea that a given query can be rewritten so that the DBMS is not expected to optimize it.
Finding a translation mechanism that guarantees the same result as the original query, while making optimizations inapplicable, is not obvious.
Our key insight is that this can be achieved by transforming a query with a \texttt{WHERE} clause, which is subject to extensive optimization by the DBMS and the basis for creating an efficient query plan, to a query that evaluates the \texttt{WHERE} clause's predicate on every record of the table, which cannot be meaningfully optimized; the number of records fetched by the first query must be equal to the number of times the \texttt{WHERE} predicate evaluates to \texttt{TRUE} for the second query. A different result indicates a bug in the DBMS.

\nparindent{}Listing~\ref{lst:illustrativeexample} illustrates the idea of our approach based on a bug that we found in \sqlite{} where an optimization caused a row to be erroneously omitted from the result set.
Starting from an initial database that contains a single record, we generate query~\circled{1} with a random \texttt{WHERE} condition \texttt{t0.c0 GLOB \textquotesingle-*\textquotesingle}.
\texttt{GLOB} is a regular expression operator, and \texttt{\textquotesingle-*\textquotesingle} a regular expression that should match a \texttt{\textquotesingle-\textquotesingle}, followed by any number of characters.
Since \texttt{WHERE} (and \texttt{JOIN}) clauses are performance-critical, they are subject to optimization by the DBMS.
In this example, \sqlite{} applies the \emph{LIKE optimization}~\cite{sqliteoptimizer} by using an index---which is an auxiliary data structure used for efficient lookups and implicitly created based on the \texttt{UNIQUE} constraint---to do a range search, allowing the execution engine to skip irrelevant records.
Unexpectedly, the optimization causes the DBMS to omit fetching the single record, even though it matches the specified regular expression.
Next, we translate the first query to query \circled{2} so that the DBMS is unlikely to optimize it, namely by moving the \texttt{WHERE} clause's predicate directly next to the \texttt{SELECT} keyword, which causes the query to evaluate the predicate on each record of the table.
We expect that the number of times the expression evaluates to \texttt{TRUE} corresponds to the actual number of records fetched by the first query.
However, in this example, the expression evaluates to \texttt{TRUE} for the single record in the database.
The DBMS could only meaningfully apply the incorrect optimization to the first query, but not to the second.
We reported this bug to the \sqlite{} developers, who quickly fixed it.

% https://www.sqlite.org/src/tktview?name=0f0428096f
\begin{figure}
\begin{lstlisting}[caption={Illustrative example where a bug in \sqlite{}'s \emph{LIKE} optimization caused a record to mistakenly be omitted.}, label=lst:illustrativeexample,escapeinside=||]
CREATE TABLE t0(c0 UNIQUE);
INSERT INTO t0 VALUES (-1);
|\circled{1}| SELECT * FROM t0 WHERE t0.c0 GLOB '-*'; -- {} |\bugsymbol|
|\circled{2}| SELECT t0.c0 GLOB '-*' FROM t0; -- {TRUE} |\oksymbol|
\end{lstlisting}
\end{figure}

\sloppy{}
\nparindent{}We implemented \approachnameshort{} in a tool called \emph{SQLancer}, which is available at \url{https://github.com/sqlancer}.\footnote{An artifact prepared for long-term archival is also available~\cite{norecartifact}.}
To demonstrate the generality of our approach, we evaluated \approachnameshort{} on four widely-used, production-level DBMS, \sqlite{}, MariaDB, \postgres{}, and CockroachDB.
As part of an extensive 5-month testing campaign, in which we sought to demonstrate the effectiveness of the approach and maximize its real-world impact, we found \nrTruePositives{} \emph{previously-unknown} bugs, many of which were serious, of which \fixed{} were subsequently fixed and \verified{} confirmed.
These comprised~\sumnorec{} optimization bugs, \sumcrash{} crash bugs, \sumdebug{} assertion failures, and \sumerror{} error bugs.
Although \sqlite{} has been extensively tested by PQS, \approachnameshort{} found more than 100 additional bugs in it, demonstrating \approachnameshort{}'s effectiveness.
The DBMS developers greatly appreciated our efforts.
For example, the \sqlite{} website describes our successful testing campaign~\cite{sqlitetesting} and mentions the following: ``\emph{Rigger's work is currently unpublished. When it is released, it could be as influential as Zalewski's invention of AFL and profile-guided fuzzing.}''
We believe that the simplicity, effectiveness, and low implementation effort of \approachnameshort{} will result in its broad adoption.
In summary, this paper contributes the following:
\begin{itemize}
	\item a new, effective testing technique for DBMS based on a novel test oracle for detecting optimization bugs called \approachnameshort{};
	\item an implementation of \approachnameshort{} in a tool called SQLancer;
	\item an extensive evaluation of \approachnameshort{}, which uncovered more than 150 new bugs in widely-used DBMS.
\end{itemize}

\section{Background}

\paragraph{Database management systems and SQL}
DBMS are based on a \emph{data model}, which abstractly describes how data is organized.
Most widely-used DBMS are based on the \emph{relational data model} proposed by Codd~\cite{Codd1970}---according to the DB-Engines Ranking~\cite{dbengines}, seven of the ten most popular DBMS are based on it.
In our work, we primarily aim to test such relational DBMS.
Structured Query Language (SQL)~\cite{Chamberlin:1974}, which is based on relational algebra~\cite{codd1972relational}, is the most commonly used language in relational DBMS to create databases, tables, insert rows, as well as manipulate and retrieve data.
Our approach is not directly applicable to NoSQL DBMS, as they often provide their own query languages or support only a SQL subset; however, it is applicable to the newer generation of NewSQL DBMS, which attempt to achieve the same scalability of NoSQL DBMS, but provide SQL as a query language~\cite{newsql}.

\paragraph{Automatic testing}
In this work, we focus on automatic testing, which is is an effective and practical way of finding bugs, although it cannot guarantee their absence~\cite{Howden:1978}.
Two components are crucial for an automatic testing approach.
First, an effective test case must stress significant portions of the system under test, to find bugs in them.
Second, a \emph{test oracle} is required that detects whether a certain test case executes as expected.
While various database generators~\citedbgenerators{} and query generators~\citequerygenerators{} have been proposed to generate effective test cases, test oracles have received less attention.
As part of this work, we propose an effective, cost-effective test oracle that allows detecting logic bugs in DBMS.

\paragraph{Optimizations in DBMS}
Decades of work have been devoted to query optimization~\cite{graefe1993query,elmasri2017fundamentals}.
Each DBMS typically provides a query optimizer that inspects a query, potentially simplifies it, and maps it efficiently to physical accesses (\ie{}, by selecting one of potentially multiple available access paths~\cite{selinger1979}).
Consider the two queries in Listing~\ref{lst:illustrativeexample}.
It is well understood that the primary performance gains of query optimizations stem from determining how the database records can be efficiently fetched.
Consequently, the query optimizer would focus its optimization effort on simplifying and creating an efficient query plan based on the \texttt{WHERE} clause in query~\circled{1}.
In query~\circled{2}, the predicate is evaluated once for every row in the result set, and thus provides limited space for meaningful optimization.
As detailed below, we utilize this observation to translate an optimized query to one that is less optimized.

\begin{figure*}[ht]
	\includegraphics[width=\textwidth]{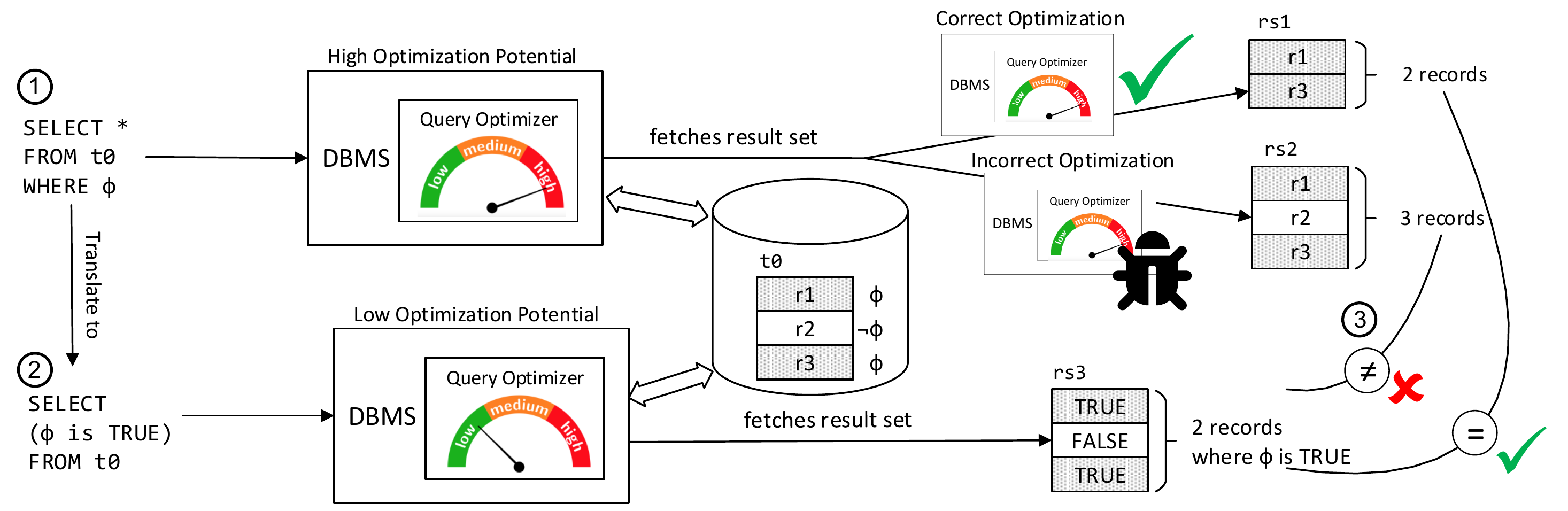}
	\caption[]{The core of the approach is the translation of an optimized query (step \circled{1}) to an unoptimized one (step \circled{2}), which allows the automatic detection of optimization bugs (step \circled{3}). \texttt{t0} is a table contained in the database, and \texttt{rs1}, \texttt{rs2}, as well as \texttt{rs3} are result sets returned by the DBMS. Predicate $\phi$ is random, but fixed.}
	\label{fig:overview}
	\reducespace{}
	
\end{figure*}

\paragraph{Differential testing}
Differential testing~\cite{mckeeman1998differential} refers to a testing technique where a single input is passed to multiple systems that are expected to produce the same output; if the systems disagree on the output, a bug in at least one of the systems has been detected.
Slutz applied this technique for testing DBMS in a system called RAGS by generating SQL queries that are sent to multiple DBMS and then observing differences in the output sets~\cite{slutz1998massive}.
While the approach was effective, the author stated that the small common core and the differences between different DBMS were a challenge.
Indeed, DBMS typically differ in the SQL dialect that they support, by deviating from the standard and providing DBMS-specific extensions~\cite{pqs}.
For example, the CockroachDB developers argued that they cannot use differential testing using PostgreSQL~\cite{Jibson2019}, which is the DBMS that is closest to it:
\emph{Correctness is difficult because we don't have any oracle of truth, which would require a known working SQL engine, which is exactly the thing we're trying to break. We are unable to use Postgres as an oracle because CockroachDB has slightly different semantics and SQL support, and generating queries that execute identically on both is tricky and doesn’t allow us to use the full CockroachDB grammar.}
In this paper, we propose an approach that allows building such a ``known working SQL engine'', namely one that is expected to be free of optimization bugs.
As argued next, it is unclear how differential testing could be used to achieve this.

\paragraph{Controlling optimizations in DBMS}
One obvious, but infeasible approach to finding optimization bugs is to realize differential testing by executing a SQL query once by disabling, and once by enabling optimizations in a DBMS to detect bug-induced deviations in the result set.
This technique has been applied on compilers~\cite{Le2014,Yang2011}, where programs were compiled without and with optimization flags.
For DBMS, the majority of optimizations cannot be disabled.
DBMS typically provide \emph{some} run-time and compile-time options to control the behavior of operators and optimizations, but these are typically very limited.
For example, the \texttt{LIKE} optimization applied to the query in Listing~\ref{lst:illustrativeexample} cannot be disabled; \sqlite{} provides only an option to control whether the operator should ignore the casing of the string. 
Similarly, some DBMS allow the specification of hints to the query optimizer for a given query~\cite{bruno2009}, which also does not apply to many optimizations.
Although modifying the DBMS to provide configuration options for all optimizations would be a possibility, doing so would require DBMS-specific knowledge and would involve a high implementation effort.

\section{Approach}
To find optimization bugs in DBMS, we propose \approachnameshort{}.
Our core insight is that a given query that is optimized by the DBMS can be transformed to another query that cannot be effectively optimized.
For brevity, we refer to the query that is potentially optimized by the DBMS as the \emph{optimized query}, and the query that is not or less optimized as the \emph{unoptimized query}.
While our translation step cannot guarantee the absence of optimizations, we found that this technique is widely applicable to disable them in practice.

\subsection{Approach Overview}
Figure~\ref{fig:overview} illustrates our approach.
In step~\circled{1}, we randomly generate an optimized query of the form \texttt{SELECT * FROM t0 WHERE $\phi$}.
Since most optimizations apply to data filtering, such as expressed in the query's \texttt{WHERE} clause, we expect that the randomly-generated query might be optimized by the DBMS.
In the figure, the database contains a single table \texttt{t0} holding the records \texttt{r1}, \texttt{r2}, and \texttt{r3}.
Assuming that the DBMS functions correctly, the result set should correspond to \texttt{rs1}, which comprises two records \texttt{r1} and \texttt{r3}.
Due to an incorrect optimization, however, it might occur that a record is omitted, or a record is mistakenly fetched.
In the example, \texttt{rs2} mistakenly additionally contains \texttt{r2}.

\nparindent{}In step~\circled{2}, we translate the optimized query to an unoptimized query of the form \texttt{SELECT ($\phi$ IS TRUE) FROM t0}.
This query lacks a \texttt{WHERE} condition.
Thus, the DBMS must fetch every record in the selected table, which effectively disables most of the optimizations that the DBMS could apply.
Furthermore, this query evaluates \texttt{$\phi{}$} as a boolean predicate on every record in the table.
This predicate should evaluate to \texttt{TRUE} for every record that is contained in the result set of the optimized query (\ie{}, for which the \texttt{WHERE} clause evaluates to \texttt{TRUE}), because the predicate must consistently yield the same value, regardless of where it is used.
The result set thus must contain two records with \texttt{TRUE}, and one record with \texttt{FALSE}.

\nparindent{}In step~\circled{3}, we pass both queries to the DBMS and compare the two result sets (i.e., \texttt{rs1}/\texttt{rs2} with \texttt{rs3}).
For the optimized query, we count the number of records, that is, $|\texttt{rs1}|=2$ for the correct execution and $|\texttt{rs2}|=3$ for the incorrect execution.
For the unoptimized query, we count the number of \texttt{TRUE} values in the result set, that is, $|\sigma_{\mathit{column_1} = \mathit{TRUE}}(r2)=2|$, which should correspond to the number of records that are fetched for the optimized query.
Since 2$\neq$3 for the incorrect execution, \approachnameshort{} detects a bug in the query optimizer.
Note that we consider only the cardinalities of the result set for the optimized query and for how many rows the expression evaluates to \texttt{TRUE} for the unoptimized query to validate the DBMS.
Our empirical evidence demonstrates that this suffices to find all optimization bugs.
For completeness, Section~\ref{sec:detercount} describes how \approachnameshort{} can be extended to also validate the records' contents.

\subsection{Translating the Query}
\label{sec:translation}
Translating an optimized query to an unoptimized one is a straightforward, automatic procedure.
As illustrated in Figure~\ref{fig:overview} step~\circled{2}, it requires moving the condition in a \texttt{WHERE} clause to after the select statement, so that it is executed on every row in the table.
As detailed next, the basic approach can be extended to cover additional features of the DBMS.

\paragraph{Multiple tables}
In a \texttt{FROM} clause, multiple tables can be specified from which records are fetched, which are typically joined by a predicate in the \texttt{WHERE} clause.
Although the previous examples only referred to a single table, our approach directly applies to multiple tables without any modifications.

\paragraph{Join clauses}
Besides \texttt{WHERE} clauses, also \texttt{JOIN} clauses can be used to join two tables.
For example, consider query~\circled{1} in Listing~\ref{lst:joinclause}, which shows an example with one (inner) \texttt{JOIN} and one \texttt{LEFT JOIN}.
The \texttt{ON} clause for inner \texttt{JOIN}s specifies that only those records should be fetched for which the condition evaluates to \texttt{TRUE} for records in both tables (\ie{}, as if the predicate would have been specified in a \texttt{WHERE} clause).
A \texttt{LEFT JOIN} fetches all records that an inner \texttt{JOIN} fetches; in addition, it fetches all records from the left table that do not have a matching record in the right table, by assuming selected columns from the right table to be \texttt{NULL}.
These, and the other types of joins (\eg{}, \texttt{NATURAL JOIN}s, \texttt{RIGHT JOIN}s, and \texttt{FULL JOIN}s) can be left unmodified during translation.
Query~\circled{2} shows that only the \texttt{WHERE} condition \texttt{ t2.c0 = 5} was moved after the \texttt{SELECT} clause, and that the \texttt{JOIN}s were copied.
An alternative strategy that could find additional bugs in joins would be to move their \texttt{ON} clauses as well, which would require translating them to multiple unoptimized queries (see Section~\ref{sec:discussion}).

\begin{figure}
\begin{lstlisting}[caption={Join clauses can be copied during translation.}, label=lst:joinclause,escapeinside=||]
|\circled{1}| SELECT * FROM t0 LEFT JOIN t1 ON t0.c0 = t1.c0 JOIN t2 ON t2.c0 > t0.c1 WHERE t2.c0 = 5;
|\circled{2}| SELECT ((t2.c0 = 5) IS TRUE) FROM t0 LEFT JOIN t1 ON t0.c0 = t1.c0 JOIN t2 ON t2.c0 > t0.c1;
\end{lstlisting}
\reducespace{}
\end{figure}

\paragraph{ORDER BY}
\texttt{ORDER BY} clauses do not influence the cardinality of the result set.
Thus, the unoptimized query can either omit or replace it during translation, to test for bugs related to this feature.

\paragraph{GROUP BY}
\texttt{GROUP BY} clauses group records with same values and are often used in combination with aggregate and window functions.
These clauses, if present in the optimized query, can be copied to the unoptimized query.
If so, an additional query is required to sum up the intermediate counts of the individual groups, assuming that an aggregate function is used to sum up the records for which the expression evaluates to \texttt{TRUE} (\cf{}~Section~\ref{sec:detercount}).

\subsection{Determining the Row Count}
\label{sec:detercount}
Figure~\ref{fig:overview}, step~\circled{3} does not illustrate how to compute the counts for the optimized and unoptimized queries. 
We apply different strategies for this.
The naive approach is to iterate through the result set to determine the count, which is applicable for both queries.
The second, more efficient strategy---the performance gain varies on various parameters, such as the number of rows in the database---relies on aggregate functions provided by the DBMS to retrieve the count, but might result in bugs being overlooked, since the increased complexity of the query might make optimizations inapplicable.
To balance performance and bug-finding capabilities, we alternate between both strategies.

\paragraph{Optimized query.} Listing~\ref{lst:optimizablequerycount} demonstrates the two ways how we compute the count for the optimized query from Figure~\ref{fig:overview}.
Query \circled{1} represents the naive approach.
For this query, the DBMS returns a result set \texttt{rs}, through which SQLancer iterates to determine the count.
Query \circled{2} uses \texttt{COUNT(*)} to count the number of records by relying on the DBMS for this.
This is more efficient because the DBMS might optimize the query, and also because the overhead for crossing the boundaries between the DBMS and SQLancer is avoided~\cite{dbmsjit}.
SQLancer only needs to retrieve the count from the single record in the result set \texttt{rs} returned by the DBMS.

\begin{figure}
\begin{lstlisting}[caption={We alternate between two strategies for determining the count for the optimized query.}, label=lst:optimizablequerycount,escapeinside=||]
|\circled{1}| SELECT * FROM t0 WHERE |$\phi$| -- while (rs.next()) count++
|\circled{2}| SELECT COUNT(*) FROM t0 WHERE |$\phi$|  -- count=rs.getInt(1)
\end{lstlisting}
\reducespace{}
\end{figure}

\paragraph{Unoptimized query.}
For the unoptimized query, we assume that since the DBMS is unable to optimize the query, it is unable to optimize an aggregate function applied to it as well.
Since using an aggregate function is more efficient, we use only this strategy (see Listing~\ref{lst:unoptimizedcount}).
The \texttt{SUM()} function adds up the predicate values by interpreting \texttt{TRUE} as one, and \texttt{FALSE} as well as \texttt{NULL} as zero.
DBMS such as PostgreSQL and CockroachDB do not provide implicit conversions from booleans to integers, and require an additional cast.
 
\begin{figure}
\begin{lstlisting}[caption={We use an aggregate function to determine the count for the unoptimized query.}, label=lst:unoptimizedcount,escapeinside=||]
SELECT SUM(count) FROM (SELECT |$\phi$| IS TRUE) as count FROM t0 -- count = rs.getInt(1)
\end{lstlisting}
\reducespace{}
\end{figure}

\paragraph{Records content}
Our basic idea can be extended to check the records' contents.
To this end, the query generated by step~\circled{2} must list each column in addition to the predicate.
The records for which the predicate evaluates to \texttt{TRUE} for the unoptimized query can then be compared with those fetched for the optimized query in step~\circled{3}.
However, retrieving and comparing the result sets makes it necessary to use the slower naive strategy presented above.
Checking the records' contents allowed us to find an additional bug in an \sqlite{} extension, albeit not in its optimizer.
We speculate that doing so was not more effective because we are unaware of any optimizations that transform the fetched content.
Furthermore, while it is possible that a DBMS returns an incorrect result set with the correct cardinality, our empirical evidence suggests that such bugs are unlikely to occur.

\subsection{Corner Cases and Limitations}
\label{sec:limitations}
We tested a large subset of each DBMS' functionality and, in the process, identified general limitations as well as three \sqlite{} corner cases that need to be specially treated by our approach.
We do not consider these limitations to be essential, as they did not hinder the approach in finding bugs.

\paragraph{Ambiguous queries}
SQL queries can be ambiguous, and thus it might be possible that a DBMS returns a different result for the optimized query than the unoptimized one, which was also a challenge for previous work~\cite{apollo}.
In practice, we found subqueries to be problematic, especially when comparing the result of a subquery that might return more than one record with a value.
Thus, we decided to disable the generation of subqueries and will consider generating unambiguous subqueries as part of future work.

\paragraph{Nondeterministic functions}
A query might be unambiguous, but yield a different result between the optimized and unoptimized queries due to nondeterministic functions.
Such functions include random number generators and those that return the current time.
To prevent false positives, we disabled their generation.

\paragraph{Short circuit evaluation.}
% https://www.sqlite.org/src/tktview?name=51f5b6e427
Our approach is not applicable to detect bugs where an optimization results in an exception or error being ``optimized way''.
This is due to SQL not specifying whether the \texttt{AND} and \texttt{OR} operators must short-circuit.
We found that DBMS can handle this inconsistently between the optimized and unoptimized query.
Consider a predicate $\phi_{ok}$ \texttt{AND} $\phi{}_{err}$, where $\phi{}_{err}$ results in an error when executed.
If $\phi_{ok}$ is executed first and yields \texttt{FALSE}, the DBMS might avoid also evaluating $\phi{}_{err}$, causing the statement to execute without errors.
Otherwise, the execution of $\phi{}_{err}$ results in an error.
Consequently, our approach cannot detect incorrect optimizations that prevent expected errors to occur.

\sloppy{}
\paragraph{Other features.}
Our approach does neither directly apply to \texttt{DISTINCT} clauses nor to queries that compute results over multiple records such as aggregate as well as window functions, which is also a limitation that affects PQS.
Also these features are optimized, meaning that their implementation might be affected by optimization bugs as well.
We believe that our high level idea of translating an optimized to an unoptimized query could also be extended to be applicable in this context.

\paragraph{Number comparisons in \sqlite{}}
One of the three \sqlite{} corner cases that caused problems was that SQLite3 considered floating-point numbers and integers that represent the same value to be equal, also when using the \texttt{DISTINCT} keyword, which caused inconsistent results.
In Listing~\ref{lst:distinctsqlite3}, the \texttt{DISTINCT} keyword in the view \texttt{v0} resulted in only one of the records being fetched---which one is unspecified and differed between the optimized and unoptimized query.
For query~\circled{1}, \texttt{0} was fetched; thus, the string concatenation yielded the value \texttt{00.1}, which evaluated to \texttt{TRUE}.
For query~\circled{2}, \texttt{0.0} was fetched from the view, which resulted in the concatenated string \texttt{0.00.1}, which evaluated to \texttt{FALSE}.
Since such false positives were uncommon in SQLite3, and not present in the other DBMS, we initially manually filtered out such false positives, but then introduced an option to avoid generating \texttt{DISTINCT} keywords in views.

% https://www.sqlite.org/src/tktview?name=c4130c33be
\begin{lstlisting}[caption={The \texttt{DISTINCT} keyword in views can cause inconsistent results in \sqlite{}.}, label=lst:distinctsqlite3, escapeinside=||]
CREATE TABLE t0(c0);
INSERT INTO t0(c0) VALUES (0.0), (0);
CREATE VIEW v0(c0) AS SELECT DISTINCT c0 FROM t0;
|\circled{1}| SELECT COUNT(*) FROM v0 WHERE v0.c0 || 0.1; -- 1
|\circled{2}| SELECT (v0.c0 || 0.1) IS TRUE FROM v0; -- 0
\end{lstlisting}

\paragraph{Input columns in \sqlite{}}
The second \sqlite{} corner case concerned the \texttt{dbstat} extension in SQLite (see Listing~\ref{lst:inputcolumn}).
The \texttt{WHERE} clause \texttt{stat.aggregate = 1} set an configuration option to \texttt{TRUE}, which changed the behavior of the query and causes a record to be fetched.
When we used this predicate directly after the \texttt{SELECT} clause, however, the column was not used as an input and no record was fetched.
We addressed this by avoiding the generation of clauses that set the configuration option for this specific column and extension.

\begin{lstlisting}[caption={Input columns in \sqlite{} can change the behavior of queries}, label=lst:inputcolumn]
CREATE VIRTUAL TABLE stat USING dbstat;
SELECT * FROM stat WHERE stat.aggregate = 1; -- fetches one record
SELECT stat.aggregate = 1 FROM stat; -- FALSE
\end{lstlisting}

\paragraph{Ambiguous \texttt{GROUP BY}s in \sqlite{}}
The third \sqlite{} corner case was given by ambiguous \texttt{GROUP BY}s in a view, which caused problems in combination with other features (such as optimizer hints, see Listing~\ref{lst:ambiggroupby}).
All other DBMS that we tested prohibited such ambiguous \texttt{GROUP BY}s and returned an error for the view creation.
In our complete testing period, we encountered such cases seldom, which is why we did not address this in SQLancer.

\begin{lstlisting}[caption={Ambiguous \texttt{GROUP BY}s in \sqlite{} can cause inconsistent results}, label=lst:ambiggroupby]
CREATE TABLE t0(c0, c1, c2, PRIMARY KEY(c2)) WITHOUT ROWID;
CREATE INDEX i0 ON t0(CAST(c1 AS INT));
CREATE VIEW v0 AS SELECT 0, c0 FROM t0 GROUP BY 1 HAVING c2;
INSERT INTO t0(c2) VALUES('');
INSERT INTO t0(c1, c2) VALUES(1, 1);
SELECT * FROM v0 WHERE UNLIKELY(1); -- {}
SELECT UNLIKELY(1) FROM v0; -- TRUE
\end{lstlisting}

\subsection{Query and Database Generation}
\label{sec:generation}
The random and targeted generation of databases~\citedbgenerators{} and queries~\citequerygenerators{} for different workloads and purposes has been widely explored, and is not a contribution of this paper.
Our approach can be applied based on any randomly-generated or existing database.
It can also be applied to any random query generator that prevents the generation, or ignores errors in the corner cases described in Section~\ref{sec:limitations}.
Thus, we explain our database and query generator only for completeness.

\nparindent{}In our work, we base the generation of databases and queries on SQLancer~\cite{pqs}, which we extended to cover additional DBMS (\ie{} CockroachDB and MariaDB), as well as SQL features (\eg{} additional data types, operators, and functions).
SQLancer generates a database by randomly creating tables, indexes, inserting data, as well as by updating and deleting data to stress the DBMS in an attempt to increase the chances of finding bugs.
The core part of SQLancer's random query generation is the generation of random expressions, which we use in \texttt{WHERE} and \texttt{JOIN} clauses.
SQLancer generates these expressions heuristically, by selecting one of the applicable options.
The applicable options are specific to a given DBMS, since DBMS vary in which operators they support and which implicit conversions they perform.
The generation of the expressions is based on the grammar of the respective DBMS and the schema of the current database (to generate valid references to columns and tables).

\section{Evaluation}
The goal of our evaluation was to demonstrate the effectiveness and generality of our approach.
To this end, we tested \approachnameshort{} on four widely-used DBMS: \sqlite{}, MariaDB, PostgreSQL, and CockroachDB.
As part of this, we extended SQLancer by a database and query generator for MariaDB as well as CockroachDB, and enhanced these components for \sqlite{} and \postgres{} (see Section~\ref{sec:generation}).
To maximize our real-world impact, we invested significant time and effort over a five-month period, which allowed us to find \nrTruePositives{} true, previously unknown bugs.
Furthermore, we analyzed the bug reports in order to obtain a better understanding on which kinds of bugs \approachnameshort{} can find.
Since PQS is the closest-related work, we compared PQS with \approachnameshort{}.

\subsection{Methodology}

\paragraph{Tested DBMS}
We focused on testing four important, widely-used DBMS: \sqlite{}, \postgres{}, MariaDB, and CockroachDB (see Table~\ref{tbl:dbmsoverview}).
According to the DB-Engines Ranking~\cite{dbengines}, the Stack Overflow's annual Developer Survey~\cite{stackoverflowsurvey}, and GitHub, these DBMS are among the most popular and widely-used DBMS.
\sqlite{} is the most widely deployed DBMS overall, used in most major web browsers, mobile phones, and embedded systems.
The authors of \sqlite{} speculate that over one trillion \sqlite{} databases are in active use~\cite{sqlitemostdeployed}.
MySQL ranks on top of most popularity rankings.
However, MySQL's binaries and its source code is provided only for release versions, which are typically published every 2-3 months, which makes it tedious to filter out test cases that trigger the same underlying bug, as also noted previously~\cite{pqs}.
Furthermore, only some of the bugs found by PQS have been fixed, providing fewer incentives to test this DBMS.
Thus, we decided to test MariaDB, which is a fork of MySQL, and uses an open-source development process.
Since MariaDB shares much code with MySQL, we believe that the results are similar to those that would be obtained when testing MySQL.
PostgreSQL is also a popular DBMS; it seems to be more robust than most other DBMS, and the PQS work could find only a single logic bug in it~\cite{pqs}.
CockroachDB~\cite{cockroachdb} is a recent commercial NewSQL DBMS~\cite{newsql}.
It has received much attention and is highly popular on GitHub, although it has a low rank on the other popularity rankings.
We tested only CockroachDB's free community edition, and not the commercial enterprise edition.

\begin{table}
\tablesize{}
\caption{The DBMS we tested are popular, complex, and most have been developed for a long time.}
\label{tbl:dbmsoverview}
\begin{tabular}{>{\RaggedLeft}p{0,8cm} >{\RaggedLeft}p{1,0cm} >{\RaggedLeft}p{1,2cm} >{\RaggedLeft}p{1,2cm} >{\RaggedLeft}p{0,5cm} >{\RaggedLeft}p{1cm}} % >{\RaggedLeft}p{1,2cm} 
\toprule{}
& \multicolumn{3}{c}{Popularity Rank} & & \\
\cmidrule(r){2-4}
DBMS & DB-Engines & Stack Overflow & GitHub Stars & LOC & First Release\\ % & Age \\
\midrule
\sqlite{} & 11 & 4 & 1.5k & 0.3M & 2000 \\% & 20 years \\ % 306,754
MariaDB & 13 & 7 & 3.2k & 3.6M  & 2009\\% & 11 years\\ % 3459934
\postgres{} & 4 & 2 & 6.3k & 1.4M & 1996\\ % & 24 years \\ % 1,418,009
CockroachDB & 75 & -& 17.7k & 1.1M  & 2015\\% & 5 years\\
\bottomrule
\end{tabular}
%\reducespace{}

\end{table}

\paragraph{Testing focus}
The developer's reaction times was a significant factor that determined on which DBMS we concentrated our testing efforts.
The \sqlite{} developers were most responsive in fixing bugs; typically, they would fix a bug within hours of us reporting it.
Thus, we invested significantly more time into testing \sqlite{} than for the other DBMS.
Besides testing \sqlite{}'s core, we tested three important extensions that are included as part of \sqlite{}'s source code, but need to be enabled during build time.
One extension enables Full Text Search (FTS) for \sqlite{}, which was subject to extensive investigations by security researchers~\cite{magellan}, as it is, for example, enabled in Google Chrome.
R-Tree is an important index structure for spatial objects that is designed for efficiently supporting range queries~\cite{rtree}.
\texttt{DBSTAT} is a \emph{virtual table} that allows querying information about the content of a \sqlite{} database. % https://www.sqlite.org/dbstat.html
We also invested significant effort into testing \postgres{}; however, we were unable to find any interesting bugs, which is why the developer's reaction time was insignificant.
The CockroachDB developers quickly confirmed our bugs, and fixed many of them within days, especially those in the query optimizer.
The MariaDB developers quickly confirmed our bugs; however, only one was fixed.
We stopped testing MariaDB after reporting the initial bugs, due to the difficulty of filtering out duplicates.

\paragraph{Existing testing efforts}
All DBMS are extensively tested, which we want to illustrate based on SQLite, and CockroachDB, which both documented their testing efforts.
\sqlite{} likely has the most impressive testing effort, which is documented on the SQLite homepage~\cite{sqlitetesting}.
The \sqlite{} developers follow a design process inspired by the DO-178B guidelines~\cite{richardhippinterview}, which are concerned with the safety of safety-critical software used in certain airborne systems.
They achieve 100\% modified condition/decision coverage~\cite{richardhippinterview}, which implies that every branch has been taken and falls through at least once.
Besides, they employ out-of-memory testing, I/O error testing, crash testing, compound failure testing, fuzz testing, and dynamic analysis~\cite{sqlitetesting}.
We believe that CockroachDB is an interesting target, because the developers have put significant effort into developing and using automatic testing techniques, which they run as part of their continuous integration.
For example, they have been running a grammar-based fuzzer on CockroachDB that has found 70 bugs such as crashes and hangs within 3 years~\cite{Jibson2016, Jibson2019}. % https://www.cockroachlabs.com/blog/testing-random-valid-sql-in-cockroachdb/
In addition, they ported SQLSmith to Go to use it as an alternative query generator, which found over 40 bugs~\cite{Jibson2019}.
However, as noted by them~\cite{Jibson2019}, ``[t]his kind of fuzz testing is not able to deduce correctness'', which is the gap that we are filling with \approachnameshort{}. 
When we reported bugs, they actively worked on enhancing their testing infrastructure to find similar correctness bugs based on domain knowledge and random testing techniques.

\paragraph{Testing methodology}
We implemented our approach iteratively and applied it to the DBMS under test after each iteration.
Typically, we added a new feature to the random query generator (\eg{}, a new operator) or database generator (\eg{}, a new data type), after which we continued testing the DBMS.
While some bugs were found seconds after implementing a feature, others were found only after weeks.
After finding a bug, we reduced the test case.
Although special query reducers have been proposed~\cite{slutz1998massive,apollo}, we found that C-Reduce~\cite{creduce}, a tool that was originally developed for reducing C/C++ programs, was sufficient for our use case.
We also manually reduced and canonicalized test cases, to reduce the developer's debugging effort.
After excluding potential duplicates, we reported issues on the bug tracker, mailing list, or via a private report (when we considered the bug to potentially be a security issue).
We did not analyze any potential security issues, since the focus of this work are optimization bugs.
Until a bug was fixed, we tried to avoid generating patterns that triggered the bug.
We invested significant time and effort in testing, as well as in triaging and reporting bugs, and opened a total of \nrReported{} bug reports.
Due to the iterative implementation and deployment of our tool, we cannot provide any detailed statistics on the total run time or efficiency.

\subsection{Selected Bugs}
Next, we show a selection of bugs found by \approachnameshort{} to give an intuition on what kind of \emph{interesting} bugs it can find.
These bugs are necessarily biased---we found many less interesting bugs, but also other interesting ones that we had to omit.
The full list of bugs can be found as part of the supplementary material supplied with the manuscript.
For brevity, we omit the unoptimized query where it can be directly derived from the optimized one.
Rather, we highlight the actual, incorrect result with a \bugsymbol{} symbol, and the expected result with a \oksymbol{} symbol.

\subsubsection{Selected \sqlite{} Bugs}

\paragraph{Incorrect \texttt{IN} optimization.}
The SQL \texttt{IN} operator allows checking whether a value on the left side is contained in a set of values on the right side.
Previously, \sqlite{} implemented an optimization that transformed an expression of the form \texttt{X IN (Y)} to \texttt{X=Y} (note that Y is a single value).
For the \texttt{=} operator, \sqlite{} performs implicit type conversions based on the \emph{affinity} of an operand (\eg{} the column type), while it is not supposed to perform them for the \texttt{IN} operator.
We thus found that this optimization is incorrect in the presence of affinity conversions (see Listing~\ref{lst:inoperatoropt}), as it caused \sqlite{} to mistakenly convert the string \texttt{`1'} to an integer in the query, thus unexpectedly fetching a record.
We found other, similar bugs related to affinity conversions (\eg{}, a bug in the constant propagation).
We believe that affinity conversions are difficult to reason about, and our findings seem to demonstrate that this mechanism is error prone.

\begin{figure}
\begin{lstlisting}[caption={A bug in the \texttt{IN} affected expressions with affinity.},label=lst:inoperatoropt, escapeinside=||]
CREATE TABLE t0(c0 INT UNIQUE);
INSERT INTO t0(c0) VALUES (1);
SELECT * FROM t0 WHERE '1' IN (t0.c0); -- {1} |\bugsymbol| {} |\oksymbol|
\end{lstlisting}
\reducespace{}
\end{figure}

\begin{figure}
% https://www.sqlite.org/src/tktview?name=767a8cbc6d
\begin{lstlisting}[caption={Commuting the operator resulted in the partial index being mistakenly used.},label=lst:collate, escapeinside=!!]
CREATE TABLE t0(c0 COLLATE NOCASE, c1);
CREATE INDEX i0 ON t0(0) WHERE c0 >= c1;
INSERT INTO t0 VALUES('a', 'B');
SELECT * FROM t0 WHERE t0.c1 <= t0.c0; -- {}!\bugsymbol!, {a|B}!\oksymbol!
\end{lstlisting}
\reducespace{}
\end{figure}

\paragraph{Operator commuting disregards \texttt{COLLATE}}
We found a bug where commuting an operator mistakenly resulted in matching an inapplicable partial index (see Listing~\ref{lst:collate}).
The \texttt{COLLATE NOCASE} clause specifies that when this column is used in string comparisons, the casing of strings should be disregarded; however, since the \texttt{c1} column is used on the left hand side and lacks a \texttt{COLLATE}, the casing is assumed to be relevant.
Thus, the lowercase characters \texttt{`a'} is assumed to be greater than the uppercase \texttt{`B'}, making the predicate yield \texttt{TRUE}.
However, the record was not fetched.
The cause for the bug was that \sqlite{} commuted the operator, while updating the expressions' \texttt{COLLATE}s, which subsequently caused it to match the partial index, since insufficient information was preserved to verify whether the expression correctly qualifies the index.
The bug was fixed by adding logic to maintain this information.

\subsubsection{Selected CockroachDB Bugs}

\paragraph{Vectorization engine bug.}
We found \cockroachdbVECTORIZE{} bugs related to CockroachDB's vectorization engine, one of which is illustrated by Listing~\ref{lst:vectorization}.
The query is expected to fetch two records, because the two tables are joined without filtering out records, effectively computing the table's cross product (\texttt{|t0|*|t1| = 2*1 = 2}).
However, only one record was fetched.
The core reason for this bug was that the empty set of equality columns for the join between the two tables was handled incorrectly for hash joins by the vectorized execution engine.

\begin{figure}
% https://github.com/cockroachdb/cockroach/issues/44207
\begin{lstlisting}[caption={A bug in the vectorization engine caused a record to be omitted.}, label=lst:vectorization, escapeinside=!!]
SET SESSION VECTORIZE=experimental_on;
CREATE TABLE t1(c0 INT);
CREATE TABLE t0(c0 INT UNIQUE);
INSERT INTO t1(c0) VALUES (0);
INSERT INTO t0(c0) VALUES (NULL), (NULL);
SELECT * FROM t0, t1 WHERE t0.c0 IS NULL; -- {NULL|0}!\bugsymbol{}! {NULL|0, NULL|0}!\oksymbol{}!
\end{lstlisting}
\reducespace{}
\end{figure}

\paragraph{Incorrect handling of filters}
We found a bug that exposed that CockroachDB, in rare cases, incorrectly handled \texttt{CHECK} constraints, which are used to refine the ranges in filters, causing the query in Listing~\ref{lst:filters} to incorrectly fetch a record, even though the predicate should evaluate to \texttt{NULL}.
While this specific bug was fixed by introducing a missing normalization rule, the underlying cause for the bug was fixed in a subsequent commit that extended and refactored CockroachDB's index constraints library.

\begin{figure}
\begin{lstlisting}[caption={A bug in the handling of filters unexpectedly caused a record to be fetched.}, label=lst:filters, escapeinside=||]
CREATE TABLE t0(c0 BOOL UNIQUE, c1 BOOL CHECK (true));
INSERT INTO t0(c0) VALUES (true);
SELECT * FROM t0 WHERE t0.c0 AND (false NOT BETWEEN SYMMETRIC t0.c0 AND NULL AND true); -- {TRUE}|\bugsymbol{}| {}|\oksymbol{}|
\end{lstlisting}
\reducespace{}
\end{figure}

\subsubsection{Selected MariaDB Bugs}

\paragraph{Incorrect string range scan.}
We found a bug in MariaDB where a range scan on an index was applied incorrectly for a string comparison (see Listing~\ref{lst:rangescan}).
As explained by the MariaDB developers, the optimizer incorrectly constructed the range \texttt{NULL < x <= 0} for the range scan, even though the upper limit should be \texttt{2}, leading to only one row being fetched.
The reason for that was that space characters like \texttt{\textbackslash{}n} were handled inconsistently.
This bug also affected MySQL, and was the only bug that we reported that was fixed by the MariaDB developers.

\begin{figure}
\begin{lstlisting}[caption={A bug in the range scan resulted in a row being omitted.}, label=lst:rangescan, escapeinside=||]
CREATE TABLE t0(c0 INT UNIQUE);
INSERT INTO t0 VALUES(NULL),(NULL),(NULL),(NULL),(1),(0);
SELECT * FROM t0 WHERE c0 < '\n2'; -- {0}|\bugsymbol{}| {0, 1}|\oksymbol{}|
\end{lstlisting}
\reducespace{}
\end{figure}

\paragraph{Incorrect number comparison.}
We found a bug where a comparison of a floating-point number with an integer column, on which an index is created, yielded an incorrect result (see Listing~\ref{lst:floatint}).
The comparison \texttt{0.5=c0} should evaluate to \texttt{FALSE}, because \texttt{c0} should, after an implicit type conversion, evaluate to the value \texttt{1.0}, and \texttt{0.5$\neq$1.0}.
However, the query unexpectedly fetched the single record stored in the table.
While this bug was quickly confirmed and reproduced for both MariaDB and MySQL, it has not been addressed yet.

\begin{figure}
\begin{lstlisting}[caption={A predicate comparing a float-point number with an integer unexpectedly evaluated to \texttt{TRUE}.}, label=lst:floatint, escapeinside=||]
CREATE TABLE t0(c0 INT);
INSERT INTO t0 VALUES (1);
CREATE INDEX i0 ON t0(c0);
SELECT * FROM t0 WHERE 0.5 = c0; -- {1}|\bugsymbol{}| {}|\oksymbol{}|
\end{lstlisting}
\end{figure}

\subsection{Bugs Overview}

\paragraph{General bug statistics}
Table~\ref{tbl:nrfoundbugs} shows the number of bugs that we reported and their status.
Out of the \nrReported{} bug reports, \nrTruePositives{} turned out to be previously-unknown \emph{true bugs}.
\fixed{} of these bugs were addressed by code changes, demonstrating that the developers took our bug reports seriously.
\verified{} bugs were verified but have not been addressed yet, and \fixedindocumentation{} bug reports were addressed by documentation changes.
\nrFalsePositives{} turned out to be \emph{false bugs}.
Out of these, \closednotabug{} were not considered to be bugs, either because an internal error that we considered to be unexpected was actually expected by the developers, or because we were not yet aware of the limitations of the approach (see Section~\ref{sec:limitations}).
As we tested the latest version of each DBMS, only \closedduplicate{} bugs turned out to be already known.

\begin{table}
\caption{We found \nrTruePositives{} bugs in \sqlite{}, MariaDB, \postgres{}, and CockroachDB, \fixed{} of which have been fixed.}
\label{tbl:nrfoundbugs}
\begin{tabular}{l r r r r r}  
\toprule
 &  &  & \multicolumn{2}{c}{Closed} \\
\cmidrule(r){4-5}
DBMS & Fixed & Verified & Intended & Duplicate \\
\midrule
\sqlite{} & \sqlitefixedInDocsOrCode & \sqliteverified{} & \sqliteclosednotabug{} & \sqliteclosedduplicate{}\\  % \sqliteopen{}
MariaDB &  \mariadbfixedInDocsOrCode & \mariadbverified{} & \mariadbclosednotabug{} & \mariadbclosedduplicate{} \\  % \mysqlopen{}
\postgres{} & \postgresfixedInDocsOrCode & \postgresverified{} & \postgresclosednotabug{} & \postgresclosedduplicate{} \\ %  \postgresopen{}
CockroachDB & \cockroachdbfixedInDocsOrCode & \cockroachdbverified{} & \cockroachdbclosednotabug{} & \cockroachdbclosedduplicate{} \\ %  \cockroachdbopen{}
\bottomrule
\end{tabular}
\end{table} 

\paragraph{Oracles}
Table~\ref{tbl:oracles} shows the oracles that we used to find the bugs.
We found \sumnorec{} bugs with the \approachnameshort{} oracle, which was the primary focus of our work.
Besides, we identified \sumerror{} bugs through unexpected internal errors, either while creating the database, or when sending queries to it.
We found these bugs by annotating a list of expected errors to each SQL statement~\cite{pqs}.
SQLancer identified also many crash bugs, since it implicitly acts as a grammar-based fuzzer.
We built the debug versions of \sqlite{} and PostgreSQL, which allowed us to find \sumdebug{} debug assertion failures.
Furthermore, we found \sumcrash{} crash bugs---which also included hang bugs---in release builds (but not necessarily in released versions of the DBMS).

\paragraph{Additional clauses}
Section~\ref{sec:translation} mentions that optionally, \texttt{ORDER BY} and \texttt{GROUP BY} clauses can be generated, to further stress the query optimizer.
We found one logic bug, and one crash bug with an \texttt{ORDER BY} clause.
We found only one error bug using a \texttt{GROUP BY} clause.
Overall, these two clauses did not contribute much to NoREC’s bug-finding capabilities; however, their implementation requires little effort, which might still justify their implementation.

\begin{table}
\center
\caption{We found \sumnorec{} bugs with the \approachnameshort{} oracle, \sumerror{} through unexpected errors, \sumdebug{} by debug assertion failures, and \sumcrash{} by crashes that occur in release version.}
\label{tbl:oracles}
\begin{tabular}{l r r r r}
\toprule{}
 &  &  & \multicolumn{2}{c}{Crash} \\
\cmidrule(r){4-5}
DBMS & Logic & Error & Release & Debug \\
\midrule
\sqlite{} & \sqlitenorec{} & \sqliteerror{} & \sqlitecrash{} & \sqlitedebug{}\\
MariaDB & \mariadbnorec{} & \mariadberror{} & \mariadbcrash{} & \mariadbdebug{} \\
\postgres{} & \postgresnorec{} & \postgreserror{} & \postgrescrash{} & \postgresdebug{}\\
CockroachDB & \cockroachdbnorec{} & \cockroachdberror{} & \cockroachdbcrash{} & \cockroachdbdebug{}\\
\bottomrule
\end{tabular}
\end{table}

\paragraph{\sqlite{}}
We found most bugs in \sqlite{}, which is expected, since we invested most effort and energy into testing it.
Out of the \nrBugssqlite{} \sqlite{} bugs, \sqliteCORE{} affected the \sqlite{} core.
A smaller portion affected extensions; we found \sqliteRTREE{} bugs in RTREE, \sqliteFTS{} in FTS, and \sqliteDBSTAT{} in DBSTAT.
Note that some bugs that we found in these extensions affected virtual tables in general, on which these and other extensions are based.
While we were testing \sqlite{}, the developers added support for \emph{generated columns}~\cite{sqlitegeneratedcolumns}, which are columns that are computed based on other columns.
After this feature was merged to trunk, but before it was released, we found \sqliteGENERATEDCOLUMN{} bugs in it, contributing significantly to its correctness.
Besides logic bugs, \sqlitedebug{} bugs manifested themselves as debug assertion failures.
This high number can be explained by previous work such as PQS having omitted testing them.
A number of these assertions did not indicate real bugs in the \sqlite{} core; rather, they indicated the omission of corner cases in the testing logic that \sqlite{} uses~\cite{assertsqlite}.
The high number of crashes in release builds is surprising, considering that PQS found only 2 crash bugs in \sqlite{}~\cite{pqs}.
One main reason is the aforementioned generated column feature, in which we found \sqliteGeneratedColumnSegfaults{} crash bugs.
We also found one hang bug in FTS, one bug that involved a trigger, two bugs in RTREE, and two bugs in window functions.

\paragraph{PostgreSQL}
Although we invested significant effort into testing PostgreSQL, we found only \nrBugspostgres{} bugs in it.
None of these bugs was an optimization bug.
This is consistent with previous findings; for example, PQS could find only a single logic bug in this DBMS~\cite{pqs}.
We believe that one significant reason for that is that PostgreSQL is very restrictive in what input it accepts compared to the other DBMS.
Richard Hipp, the main \sqlite{} developer, also noted that PostgreSQL in particular is a high-quality DBMS, which has had few bugs and noted that one possible reason could be their very elaborate peer review process~\cite{richardhippinterview}.

\paragraph{CockroachDB}
We found \nrBugscockroachdb{} bugs in CockroachDB.
In \cockroachdbEXPERIMENTAL{} cases, our bug reports relied on experimental features.
Of these, \cockroachdbVECTORIZE{} bugs affected the vectorizer engine (see Listing~\ref{lst:vectorization}).
Out of the \cockroachdberror{} error bugs, \cockroachdbINTERNALERROR{} were due to internal errors, for which execution resulted in displaying a stack trace along with information on where to report the bug, while the server stayed responsive. 
Based on our bug reports, the CockroachDB developers actively added testing infrastructure and reviewed code to detect similar bugs, demonstrating that they took our bug reports seriously.
For example, one of the duplicate bug report was due to an open issue that acknowledged that the issue was found based on one of our bug reports.

\paragraph{MariaDB}
We found \nrBugsmariadb{} bugs in MariaDB.
All bugs were quickly confirmed by the MariaDB developers, and three of the bugs were reproduced also for MySQL.
However, only one bug was fixed within a duration of three months, which is why we stopped testing MariaDB.
Since we invested only little time in testing MariaDB, we believe that our approach could find additional bugs in it.

\subsection{Comparison to PQS}
\label{sec:pqscomparison}
PQS is the state of the art in finding logic bugs in DBMS, which is why we want to compare \approachnameshort{}'s effectiveness with it.
We expected PQS to find a broader class of bugs, because our aim for \approachnameshort{} was to find optimization bugs.
To the best of our knowledge, there is no other publicly available tool that could detect logic bugs in DBMS to which we could compare.

\paragraph{Evaluation challenges}
Fairly comparing the effectiveness between PQS and \approachnameshort{} is difficult due to various reasons.
First, the implementation effort of PQS is significant, and limited the authors to a core subset of the respective DBMS's supported SQL dialect~\cite{pqs}.
For example, in PQS, a single comparison operator alone covers already more than 200 LOC, since it needs to support comparing arbitrary data types, which also involves implicit conversions for DBMS like \sqlite{} or MySQL.
In contrast, the implementation of the \approachnameshort{} oracle consists of less than 200 LOC, and also allows finding bugs in the optimization of complex operators and functions.
Nevertheless, we believe that a fair evaluation should disregard the amount of time that was invested into implementing the respective approach.
Another challenge is that a different set of DBMS were tested.
While both PQS and \approachnameshort{} were evaluated on \sqlite{} and \postgres{}, PQS was also evaluated on MySQL, while we evaluated \approachnameshort{} on MariaDB, and CockroachDB.
Overall, this inhibited us from doing an automatic comparison.

\paragraph{Methodology}
To provide a fair comparison, we performed a manual quantitative and qualitative comparison.
We noticed that a test case for \approachnameshort{} can typically be converted directly to an equivalent PQS test case that triggers the bug, if it can be reproduced by PQS.
In fact, we can take the unmodified query with the predicate in the \texttt{WHERE} clause and check if the bug can be reproduced by selecting an applicable pivot row.
Likewise, a PQS test case can often be converted to an equivalent \approachnameshort{} test case, by performing the translation step to an unoptimized query.
Based on this observation, we manually tried to create equivalent test cases where possible, which we then evaluated on the version of the DBMS in which the bug was found.
While we cannot completely rule out misclassifications that might be due to overlooking that a bug could be reproduced by another query, we believe that the majority of cases were clear.
Note that we considered only bugs found by the \approachnameshort{} oracle---we expect that the errors and crashes can be triggered with either of the approaches.
Overall, we investigated (1) the \sumnorec{} bugs found by the \approachnameshort{} oracle to check if they could have been found by PQS as well and (2) the 61 bugs found by PQS to check whether they could have been found by \approachnameshort{}.

\paragraph{Bugs found only by \approachnameshort{}}
PQS could detect \percNoRECBugsFoundByPQS{}\% of the bugs that were found by \approachnameshort{}.
Specifically, \approachnameshort{} detected \duplicates{} bugs for which the result set unexpectedly contained or missed duplicate records.
PQS conceptually cannot detect such bugs, because it validates only whether a randomly-selected record, which is indistinguishable from other duplicate rows, is part of the result set.
\approachnameshort{} triggered \aggregate{} bugs by aggregate functions that are used for counting the rows (\ie{}, \texttt{SUM} and \texttt{COUNT}).
PQS' main oracle relies on only checking a single row at a time, which conceptually hinders PQS in detecting bugs related to these aggregate functions.
We found \incorrectlyfetched{} cases where records were mistakenly fetched.
PQS did not detect this class of bugs, because it only checks for bugs where the pivot row is mistakenly omitted from the result set.
We believe that this not a fundamental limitation of PQS, because it could be extended to also generate queries that guarantee to omit the pivot row, enabling PQS to also detect such bugs.
Taking this into account, PQS could detect \percNoRECBugsFoundByPQSDisregardingMistakenlyFetchedRows{}\% of the bugs that \approachnameshort{} found.
A caveat that was already mentioned is that some other bugs can be detected only by a close-to-complete implementation of PQS; as noted, we did not consider this limitation in this analysis, since it is arguable which cases would be deemed to involve too much of an implementation effort. 

\paragraph{Bugs found only by PQS}
In total, \approachnameshort{} could have found \percPqsBugsFoundByNoREC{}\% of the bugs that PQS found.
By far the most common kind of bugs that were detected by PQS, but not by \approachnameshort{}, stem from the incorrect implementation of operators, functions, and other features (especially affinity conversions in \sqlite{}), both in the optimized and unoptimized case.
\approachnameshort{} failed to detect \norecunknown{} such bugs.
\norecdistinct{} bugs could not be triggered by \approachnameshort{}, since they relied on a \texttt{DISTINCT} query, which was disregarded by \approachnameshort{}, but could be supported by translating it to a \texttt{GROUP BY} clause in the unoptimized case.
While PQS, like \approachnameshort{}, is in general unable to detect bugs in aggregate functions, since it operates on a single pivot row, it can do so when a table contains only a single row, which allowed it to detect \norecaggregate{} bugs.
PQS could also find \norecintersect{} bug that was exposed when using \texttt{INTERSECT}, which PQS uses to efficiently check containment, similarly to how \approachnameshort{} uses the aggregate functions for counting.
\norecleftjoin{} bug was triggered based on a predicate in an \texttt{ON} clause of a \texttt{LEFT JOIN}.
\approachnameshort{} failed to discover that bug, because it copies left joins and predicates when translating the query.
As outlined in Section~\ref{sec:discussion}, we believe that our approach might be enhanced by implementing additional translation schemes for joins.

\paragraph{Discussion}
The comparison has demonstrated that PQS can find a number of bugs that \approachnameshort{} cannot find.
Although \approachnameshort{} could find classes of bugs that PQS is unable to find as well, such as duplicate records, it is mostly restricted to finding optimization bugs.
However, PQS incurs significant implementation overhead, since every operator and function to be tested needs to be implemented, for every DBMS that should be tested.
In contrast, \approachnameshort{} relies on a straightforward translation process, and is applicable to any database and query generator that can address the limitations mentioned in Section~\ref{sec:limitations}.
Due to the low implementation effort, \approachnameshort{} has successfully found a wide range of optimization bugs, even for \sqlite{}, which has recently been comprehensively tested by PQS.
Thus, we envision that DBMS could be tested by a combination of both approaches.
PQS could be used to test many basic operators and functions, and, being an exact oracle, it can help in establishing a ground truth.
Then, \approachnameshort{} could be used to find the lingering optimization bugs in parts not comprehensively tested by PQS.

\section{Discussion}
\label{sec:discussion}

\sloppy{}
\paragraph{Reception by the DBMS developers}
Developers of multiple DBMS told us that they appreciated our bug reports.
The \sqlite{} homepage even highlights our efforts at \url{https://www.sqlite.org/testing.html}:
``\emph{One fuzzing researcher of particular note is Manuel Rigger, [...]. Most fuzzers only look for assertion faults, crashes, undefined behavior (UB), or other easily detected anomalies. Dr. Rigger's fuzzers, on the other hand, are able to find cases where SQLite computes an incorrect answer. Rigger has found many such cases. Most of these finds are fairly obscure corner cases involving type conversions and affinity transformations, and a good number of the finds are against unreleased features. Nevertheless, his finds are still important as they are real bugs, and the SQLite developers are grateful to be able to identify and fix the underlying problems. Rigger's work is currently unpublished. When it is released, it could be as influential as Zalewski's invention of AFL and profile-guided fuzzing.}''

\paragraph{Bug importance}
We believe that we found many critical bugs that are likely to affect real users.
However, we also acknowledge that many of the other bugs that we found can be triggered only by an unlikely combination of operators or features.
Even those seemingly unimportant bugs can affect users due to the widespread use of these DBMS, for which we found evidence on the \sqlite{} mailing list, where a user reported an incorrect result for a query with a complex \texttt{WHERE} predicate. %based on the code in Listing~\ref{lst:regression}.
When confronted by another \sqlite{} user, the reporter of the bug defended themselves with the following~\cite{sqliteuserbugreport}: ``\emph{I might not spell it like that myself, but a code generator would do it (and much worse!). This example was simplified from a query generated by a Django ORM queryset using \texttt{.exclude(nullable\_joined\_table\_\_column=1)}, for instance.}''
The bug had already been fixed in the latest development version, because we previously reported the same underlying bug~\cite{notnullnotnullbug}.
We found it with a query on a view and a predicate \texttt{NOTNULL NOTNULL}, which would unlikely to be written by a programmer as well.
We speculate that also other users might be affected by such seemingly-obscure bugs; however, finding the root cause in such cases is difficult, especially when the query is generated by middleware.

\paragraph{Handling of joins}
Our translation process leaves \texttt{JOIN} clauses unmodified (see Section~\ref{sec:translation}).
Although this allowed us to find some bugs in their handling, translating them to a condition that is placed after the \texttt{SELECT} as well could uncover additional bugs.
For inner joins, this is straightforward, because an \texttt{ON} predicate $\phi{}_1$ and a \texttt{WHERE} predicate $\phi{}_2$ can be translated to a predicate $\phi{}_1$ \texttt{AND} $\phi{}_2$.
For other joins, this would be more involved.
For example, for \texttt{LEFT JOIN}s, a simple \texttt{SELECT} statement would no longer sufficient, but would require combining the results of multiple queries.
We consider additional strategies in translating \texttt{JOIN} clauses as potentially useful and part of future work.

\paragraph{Code coverage and performance}
At first sight, it might be interesting to evaluate the code coverage and run-time performance that \approachnameshort{} achieves.
However, neither helps to explain the approach's effectiveness.
With respect to coverage, it was found that code coverage is not particularly useful for fuzzing DBMS, since high coverage for the core components (\eg{}, the query optimizer) can be achieved quickly~\cite{apollo}.
Furthermore, although the \sqlite{} developers achieve 100\% MC/DC coverage in their test suite, we could still find bugs in \sqlite{}.
The run-time overhead of the query generation and translation is negligible, which is why we did not measure it in detail.
The run time is dominated by the DBMS, which needs to process the generated queries, as well as by the communication with the DBMS.
The communication overhead was lower for \sqlite{} than for the other DBMS, because \sqlite{} is an \emph{embedded} DBMS, that is, it runs within the application process of SQLancer.

\paragraph{Kind of bugs}
\sumnorec{} of the \nrTruePositives{} true bugs that we found were optimization bugs.
We believe that such bugs are most severe, since they are likely to go unnoticed by developers.
Bugs caused by crashes, assertion failures, and error represent a large fraction of the overall bugs.
However, they can be found by existing approaches, such as grammar-based fuzzers.
Furthermore, such bugs typically cause the DBMS to terminate, thus signalling the user that the DBMS malfunctions for the given query.
A possible conclusion might be that our testing approach detects also errors typically found by fuzzers, and that such bugs might be more common or easier to find than logic bugs (given that we do not have the ground truth on the total number of bugs). 

\paragraph{Fully automatic approach}
We claim that NoREC is fully automatic, as it finds bugs by repeatedly generating test cases and validating their result set, without requiring any human interaction.
As discussed in Section~\ref{sec:limitations}, corner cases exist that have to be treated specially to ensure that only true bugs are reported.
When NoREC detects a bug, it is helpful to reduce the generated test case so that it is minimal.
For our evaluation, we first automatically reduced the test case, and then manually tried to reduce it further; we consider manual reduction optional, but convenient for the DBMS developers for debugging.
Before reporting bugs, we manually checked the bug tracker to lower the chances of reporting duplicate bugs.
In practice, it would be useful to suspend running NoREC after it reports a bug, and continue its execution after the bug is fixed.

\section{Related Work}

\paragraph{Differential testing of DBMS}
\emph{Differential testing} discovers bugs by executing a given input using multiple versions of a system to detect differing outputs, which indicate a bug in one of these systems.
Similarly to metamorphic testing, differential testing has proven to be effective in many domains~\cite{mckeeman1998differential,Le2014,Yang2011,Kapus2017,Brummayer2009,Pascal2012}.
For DBMS, it was first applied by Slutz in a tool \emph{RAGS} to find bugs by executing a query on multiple different DBMS and comparing their result sets~\cite{slutz1998massive}.
While the approach was highly effective, it can only be applied to a small subset of common SQL.
Gu et al. used options and hints to force the generation of different query plans, to then rank the accuracy of the optimizer based on the estimated cost for each plan~\cite{Gu2012}. 
Jinho et al. used differential testing in a system called \emph{APOLLO} to find performance regression bugs in DBMS, by executing a SQL query on an old and newer version of a DBMS~\cite{apollo}.
They found 10 previously-unknown performance regression bugs in \sqlite{} and PostgreSQL.
While our work uses metamorphic testing to find optimization bugs, conceptually, it can be interpreted as realizing differential testing by comparing the results of an optimizing, and non-optimizing version of a DBMS.

\paragraph{Other correctness oracles for testing DBMS}
Pivoted Query Synthesis (PQS) is the state of the art in testing DBMS for logic bugs and is the most closely-related work~\cite{pqs}.
It is both an automatic testing approach as well as an oracle, and is based on the idea of testing the DBMS' correctness on a randomly-selected \emph{pivot row}.
PQS has found close to 100 bugs in widely-used DBMS, demonstrating that it is highly effective.
However, due to its high implementation effort, it might be infeasible to test all of a DBMS' supported operators and functions.
\approachnameshort{} is mainly applicable to finding optimization bugs, a subcategory of logic bugs, towards which PQS is geared.
Due to its low implementation effort, it can find bugs in components for which the implementation of PQS would be too costly.
ADUSA~\cite{khalek2008} is a query-aware database generator and generates input data as well as the expected result for a query, thus also tackling the correctness oracle problem for DBMS.
It translates the schema and query to an Alloy specification, which is subsequently solved.
The approach could reproduce various known and injected bugs in MySQL, HSQLDB, and also find a new bug in Oracle Database.
We believe that the high overhead that solver-based approaches incur might inhibit such approaches from finding more DBMS bugs.

\paragraph{Random and targeted queries}
Many query generators have been proposed for purposes such as bug-finding and benchmarking.
SQLsmith is a widely-used, open-source random query generator, which has found over 100 bugs in widely-used DBMS~\cite{sqlsmith}.
Bati et al. proposed an approach based on genetic algorithms to incorporate execution feedback for generating queries~\cite{Bati2007}.
SQLFUZZ~\cite{apollo} also utilizes execution feedback and randomly generates queries using only features that are supported by all the DBMS systems they considered.
Khalek et al. proposed generating both syntactically and semantically valid queries based on a solver-backed approach~\cite{khalek2010querygen}.
The problem of generating queries that satisfy cardinality constraints on their sub-expressions was shown to be computationally hard~\cite{Bruno2006, Mishra2008}.
Consequently, a number of heuristic and approximate approaches to generating targeted queries were proposed~\cite{Bruno2006, qrelx, Mishra2008,qgen}.
All these random-query generators can be used to find bugs such as crashes and hangs in DBMS.
When paired with the test oracle proposed in this paper, they could also be used to find logic bugs.

\sloppy
\paragraph{Random and targeted databases}
Many approaches have been proposed to automatically generate databases.
Given a query and a set of constraints, QAGen~\cite{qagen,Lo2010framework} generates a database that matches the desired query results by combining traditional query processing and symbolic execution.
Reverse Query Processing takes a query and a desired result set as an input to then generate a database that could have produced the result set~\cite{reversequeryprocessing}.
As discussed above, ADUSA is a query-aware database generator~\cite{khalek2008}.
Gray et al. discussed a set of techniques utilizing parallel algorithms to quickly generate billions-record databases~\cite{gray1994}.
DGL was proposed as a domain-specific language to generate input data following various distributions and inter-table correlations based on \emph{iterators} that can be composed~\cite{nicolas2005}.
Neufield et al. proposed to generate test data for tables with constraints by deriving a generator formula that is then translated to operators to generate the test data~\cite{neufeld1993}.
An improved database generation might enable \approachnameshort{} to find additional bugs.

\paragraph{Metamorphic testing}
Our approach is based on \emph{metamorphic testing}~\cite{Chen1998}.
Metamorphic testing addresses the test oracle problem by, based on an input and output of a system, generating a new input for which the result is known.
This technique has been applied successfully in various domains~\cite{Chen2018,Sergio2018}.
Central in this approach is the metamorphic relation, which can be used to infer the expected result.
In this work, we combine a translation and counting mechanism to establish a metamorphic relation specifically geared towards detecting optimization bugs.
A limitation of metamorphic testing is that it cannot establish a ground truth.
For NoREC, this implies that both the unoptimized and optimized query can yield an incorrect result (see Section~\ref{sec:pqscomparison}).
Similarly, the optimized query could compute the correct result, rather than the unoptimized one.

\section{Conclusion}
This paper has proposed a general, highly-effective approach for detecting bugs in DBMS, which we termed \approachname{} (\approachnameshort{}).
The core insight of \approachnameshort{} is that a given \emph{optimized} query can be translated to an \emph{unoptimized} query, which enables the construction of a test oracle that can detect \emph{optimization bugs} by comparing the result sets of the two queries.
While we believe that this work provides a solid foundation for correctness testing of DBMS, the basic idea of \approachnameshort{} could be extended by using additional or alternative translation strategies for queries, for example, by translating predicates of \texttt{JOIN} clauses.
As another example, queries could also be transformed in other ways, while still being expected to generate the same results (\eg{}, by switching the operands of commutative operators).
Furthermore, the efficiency and effectiveness of \approachnameshort{} could be enhanced by pairing it with better database and query generators.

\begin{acks}
We want to thank the DBMS developers for verifying and addressing our bug reports as well as for their feedback to our work.
Furthermore, we are grateful for the feedback received by the anonymous reviewers, Martin Kersten, as well as by the members of the AST Lab at ETH Zurich.
\end{acks}

\balance
\bibliographystyle{ACM-Reference-Format}
\bibliography{bib}

%%% -*-BibTeX-*-
%%% Do NOT edit. File created by BibTeX with style
%%% ACM-Reference-Format-Journals [18-Jan-2012].

\begin{thebibliography}{61}

%%% ====================================================================
%%% NOTE TO THE USER: you can override these defaults by providing
%%% customized versions of any of these macros before the \bibliography
%%% command.  Each of them MUST provide its own final punctuation,
%%% except for \shownote{}, \showDOI{}, and \showURL{}.  The latter two
%%% do not use final punctuation, in order to avoid confusing it with
%%% the Web address.
%%%
%%% To suppress output of a particular field, define its macro to expand
%%% to an empty string, or better, \unskip, like this:
%%%
%%% \newcommand{\showDOI}[1]{\unskip}   % LaTeX syntax
%%%
%%% \def \showDOI #1{\unskip}           % plain TeX syntax
%%%
%%% ====================================================================

\ifx \showCODEN    \undefined \def \showCODEN     #1{\unskip}     \fi
\ifx \showDOI      \undefined \def \showDOI       #1{#1}\fi
\ifx \showISBNx    \undefined \def \showISBNx     #1{\unskip}     \fi
\ifx \showISBNxiii \undefined \def \showISBNxiii  #1{\unskip}     \fi
\ifx \showISSN     \undefined \def \showISSN      #1{\unskip}     \fi
\ifx \showLCCN     \undefined \def \showLCCN      #1{\unskip}     \fi
\ifx \shownote     \undefined \def \shownote      #1{#1}          \fi
\ifx \showarticletitle \undefined \def \showarticletitle #1{#1}   \fi
\ifx \showURL      \undefined \def \showURL       {\relax}        \fi
% The following commands are used for tagged output and should be
% invisible to TeX
\providecommand\bibfield[2]{#2}
\providecommand\bibinfo[2]{#2}
\providecommand\natexlab[1]{#1}
\providecommand\showeprint[2][]{arXiv:#2}

\bibitem[\protect\citeauthoryear{Abdul~Khalek and Khurshid}{Abdul~Khalek and
  Khurshid}{2010}]%
        {khalek2010querygen}
\bibfield{author}{\bibinfo{person}{Shadi Abdul~Khalek} {and}
  \bibinfo{person}{Sarfraz Khurshid}.} \bibinfo{year}{2010}\natexlab{}.
\newblock \showarticletitle{Automated SQL Query Generation for Systematic
  Testing of Database Engines}. In \bibinfo{booktitle}{\emph{Proceedings of the
  IEEE/ACM International Conference on Automated Software Engineering}}
  (Antwerp, Belgium) \emph{(\bibinfo{series}{ASE '10})}.
  \bibinfo{publisher}{ACM}, \bibinfo{address}{New York, NY, USA},
  \bibinfo{pages}{329--332}.
\newblock
\showISBNx{978-1-4503-0116-9}
\urldef\tempurl%
\url{https://doi.org/10.1145/1858996.1859063}
\showDOI{\tempurl}


\bibitem[\protect\citeauthoryear{Bati, Giakoumakis, Herbert, and Surna}{Bati
  et~al\mbox{.}}{2007}]%
        {Bati2007}
\bibfield{author}{\bibinfo{person}{Hardik Bati}, \bibinfo{person}{Leo
  Giakoumakis}, \bibinfo{person}{Steve Herbert}, {and}
  \bibinfo{person}{Aleksandras Surna}.} \bibinfo{year}{2007}\natexlab{}.
\newblock \showarticletitle{A Genetic Approach for Random Testing of Database
  Systems}. In \bibinfo{booktitle}{\emph{Proceedings of the 33rd International
  Conference on Very Large Data Bases}} (Vienna, Austria)
  \emph{(\bibinfo{series}{VLDB '07})}. \bibinfo{publisher}{VLDB Endowment},
  \bibinfo{pages}{1243--1251}.
\newblock
\showISBNx{978-1-59593-649-3}


\bibitem[\protect\citeauthoryear{Binnig, Kossmann, and Lo}{Binnig
  et~al\mbox{.}}{2007a}]%
        {reversequeryprocessing}
\bibfield{author}{\bibinfo{person}{Carsten Binnig}, \bibinfo{person}{Donald
  Kossmann}, {and} \bibinfo{person}{Eric Lo}.}
  \bibinfo{year}{2007}\natexlab{a}.
\newblock \showarticletitle{Reverse Query Processing}.
\newblock \bibinfo{journal}{\emph{Proceedings - International Conference on
  Data Engineering}}, \bibinfo{pages}{506--515}.
\newblock
\showISBNx{1-4244-0803-2}
\urldef\tempurl%
\url{https://doi.org/10.1109/ICDE.2007.367896}
\showDOI{\tempurl}


\bibitem[\protect\citeauthoryear{Binnig, Kossmann, Lo, and \"{O}zsu}{Binnig
  et~al\mbox{.}}{2007b}]%
        {qagen}
\bibfield{author}{\bibinfo{person}{Carsten Binnig}, \bibinfo{person}{Donald
  Kossmann}, \bibinfo{person}{Eric Lo}, {and} \bibinfo{person}{M.~Tamer
  \"{O}zsu}.} \bibinfo{year}{2007}\natexlab{b}.
\newblock \showarticletitle{QAGen: Generating Query-Aware Test Databases}. In
  \bibinfo{booktitle}{\emph{Proceedings of the 2007 ACM SIGMOD International
  Conference on Management of Data}} (Beijing, China)
  \emph{(\bibinfo{series}{SIGMOD ’07})}. \bibinfo{publisher}{Association for
  Computing Machinery}, \bibinfo{address}{New York, NY, USA},
  \bibinfo{pages}{341–352}.
\newblock
\showISBNx{9781595936868}
\urldef\tempurl%
\url{https://doi.org/10.1145/1247480.1247520}
\showDOI{\tempurl}


\bibitem[\protect\citeauthoryear{Bolz, Kurilova, and Tratt}{Bolz
  et~al\mbox{.}}{2016}]%
        {dbmsjit}
\bibfield{author}{\bibinfo{person}{Carl~Friedrich Bolz}, \bibinfo{person}{Darya
  Kurilova}, {and} \bibinfo{person}{Laurence Tratt}.}
  \bibinfo{year}{2016}\natexlab{}.
\newblock \showarticletitle{Making an Embedded {DBMS} JIT-friendly}. In
  \bibinfo{booktitle}{\emph{30th European Conference on Object-Oriented
  Programming, {ECOOP} 2016, July 18-22, 2016, Rome, Italy}}.
  \bibinfo{pages}{4:1--4:24}.
\newblock
\urldef\tempurl%
\url{https://doi.org/10.4230/LIPIcs.ECOOP.2016.4}
\showDOI{\tempurl}


\bibitem[\protect\citeauthoryear{Brummayer and Biere}{Brummayer and
  Biere}{2009}]%
        {Brummayer2009}
\bibfield{author}{\bibinfo{person}{Robert Brummayer} {and}
  \bibinfo{person}{Armin Biere}.} \bibinfo{year}{2009}\natexlab{}.
\newblock \showarticletitle{Fuzzing and Delta-Debugging SMT Solvers}. In
  \bibinfo{booktitle}{\emph{Proceedings of the 7th International Workshop on
  Satisfiability Modulo Theories}} (Montreal, Canada)
  \emph{(\bibinfo{series}{SMT ’09})}. \bibinfo{publisher}{Association for
  Computing Machinery}, \bibinfo{address}{New York, NY, USA},
  \bibinfo{pages}{1–5}.
\newblock
\showISBNx{9781605584843}
\urldef\tempurl%
\url{https://doi.org/10.1145/1670412.1670413}
\showDOI{\tempurl}


\bibitem[\protect\citeauthoryear{Bruno and Chaudhuri}{Bruno and
  Chaudhuri}{2005}]%
        {nicolas2005}
\bibfield{author}{\bibinfo{person}{Nicolas Bruno} {and}
  \bibinfo{person}{Surajit Chaudhuri}.} \bibinfo{year}{2005}\natexlab{}.
\newblock \showarticletitle{Flexible Database Generators}. In
  \bibinfo{booktitle}{\emph{Proceedings of the 31st International Conference on
  Very Large Data Bases}} (Trondheim, Norway) \emph{(\bibinfo{series}{VLDB
  ’05})}. \bibinfo{publisher}{VLDB Endowment}, \bibinfo{pages}{1097–1107}.
\newblock
\showISBNx{1595931546}


\bibitem[\protect\citeauthoryear{Bruno, Chaudhuri, and Ramamurthy}{Bruno
  et~al\mbox{.}}{2009}]%
        {bruno2009}
\bibfield{author}{\bibinfo{person}{Nicolas Bruno}, \bibinfo{person}{Surajit
  Chaudhuri}, {and} \bibinfo{person}{Ravi Ramamurthy}.}
  \bibinfo{year}{2009}\natexlab{}.
\newblock \showarticletitle{Power Hints for Query Optimization}. In
  \bibinfo{booktitle}{\emph{Proceedings of the 2009 IEEE International
  Conference on Data Engineering}} \emph{(\bibinfo{series}{ICDE ’09})}.
  \bibinfo{publisher}{IEEE Computer Society}, \bibinfo{address}{USA},
  \bibinfo{pages}{469–480}.
\newblock
\showISBNx{9780769535456}
\urldef\tempurl%
\url{https://doi.org/10.1109/ICDE.2009.68}
\showDOI{\tempurl}


\bibitem[\protect\citeauthoryear{Bruno, Chaudhuri, and Thomas}{Bruno
  et~al\mbox{.}}{2006}]%
        {Bruno2006}
\bibfield{author}{\bibinfo{person}{Nicolas Bruno}, \bibinfo{person}{Surajit
  Chaudhuri}, {and} \bibinfo{person}{Dilys Thomas}.}
  \bibinfo{year}{2006}\natexlab{}.
\newblock \showarticletitle{Generating Queries with Cardinality Constraints for
  DBMS Testing}.
\newblock \bibinfo{journal}{\emph{IEEE Trans. on Knowl. and Data Eng.}}
  \bibinfo{volume}{18}, \bibinfo{number}{12} (\bibinfo{date}{Dec.}
  \bibinfo{year}{2006}), \bibinfo{pages}{1721--1725}.
\newblock
\showISSN{1041-4347}
\urldef\tempurl%
\url{https://doi.org/10.1109/TKDE.2006.190}
\showDOI{\tempurl}


\bibitem[\protect\citeauthoryear{Chamberlin and Boyce}{Chamberlin and
  Boyce}{1974}]%
        {Chamberlin:1974}
\bibfield{author}{\bibinfo{person}{Donald~D. Chamberlin} {and}
  \bibinfo{person}{Raymond~F. Boyce}.} \bibinfo{year}{1974}\natexlab{}.
\newblock \showarticletitle{SEQUEL: A Structured English Query Language}. In
  \bibinfo{booktitle}{\emph{Proceedings of the 1974 ACM SIGFIDET (Now SIGMOD)
  Workshop on Data Description, Access and Control}} (Ann Arbor, Michigan)
  \emph{(\bibinfo{series}{SIGFIDET '74})}. \bibinfo{publisher}{ACM},
  \bibinfo{address}{New York, NY, USA}, \bibinfo{pages}{249--264}.
\newblock
\urldef\tempurl%
\url{https://doi.org/10.1145/800296.811515}
\showDOI{\tempurl}


\bibitem[\protect\citeauthoryear{Chen, Cheung, and Yiu}{Chen
  et~al\mbox{.}}{1998}]%
        {Chen1998}
\bibfield{author}{\bibinfo{person}{Tsong~Y Chen}, \bibinfo{person}{Shing~C
  Cheung}, {and} \bibinfo{person}{Shiu~Ming Yiu}.}
  \bibinfo{year}{1998}\natexlab{}.
\newblock \bibinfo{booktitle}{\emph{Metamorphic testing: a new approach for
  generating next test cases}}.
\newblock \bibinfo{type}{{T}echnical {R}eport}. \bibinfo{institution}{Technical
  Report HKUST-CS98-01, Department of Computer Science, Hong Kong}.
\newblock


\bibitem[\protect\citeauthoryear{Chen, Kuo, Liu, Poon, Towey, Tse, and
  Zhou}{Chen et~al\mbox{.}}{2018}]%
        {Chen2018}
\bibfield{author}{\bibinfo{person}{Tsong~Yueh Chen}, \bibinfo{person}{Fei-Ching
  Kuo}, \bibinfo{person}{Huai Liu}, \bibinfo{person}{Pak-Lok Poon},
  \bibinfo{person}{Dave Towey}, \bibinfo{person}{T.~H. Tse}, {and}
  \bibinfo{person}{Zhi~Quan Zhou}.} \bibinfo{year}{2018}\natexlab{}.
\newblock \showarticletitle{Metamorphic Testing: A Review of Challenges and
  Opportunities}.
\newblock \bibinfo{journal}{\emph{ACM Comput. Surv.}} \bibinfo{volume}{51},
  \bibinfo{number}{1}, Article \bibinfo{articleno}{Article 4}
  (\bibinfo{date}{Jan.} \bibinfo{year}{2018}), \bibinfo{numpages}{27}~pages.
\newblock
\showISSN{0360-0300}
\urldef\tempurl%
\url{https://doi.org/10.1145/3143561}
\showDOI{\tempurl}


\bibitem[\protect\citeauthoryear{Clover}{Clover}{2019}]%
        {sqliteuserbugreport}
\bibfield{author}{\bibinfo{person}{And Clover}.}
  \bibinfo{year}{2019}\natexlab{}.
\newblock \bibinfo{title}{Bug submission: left join filter on negated
  expression including NOTNULL}.
\newblock
\newblock
\urldef\tempurl%
\url{https://www.mail-archive.com/sqlite-users@mailinglists.sqlite.org/msg117434.html}
\showURL{%
\tempurl}


\bibitem[\protect\citeauthoryear{Codd}{Codd}{1970}]%
        {Codd1970}
\bibfield{author}{\bibinfo{person}{E.~F. Codd}.}
  \bibinfo{year}{1970}\natexlab{}.
\newblock \showarticletitle{A Relational Model of Data for Large Shared Data
  Banks}.
\newblock \bibinfo{journal}{\emph{Commun. ACM}} \bibinfo{volume}{13},
  \bibinfo{number}{6} (\bibinfo{date}{June} \bibinfo{year}{1970}),
  \bibinfo{pages}{377--387}.
\newblock
\showISSN{0001-0782}
\urldef\tempurl%
\url{https://doi.org/10.1145/362384.362685}
\showDOI{\tempurl}


\bibitem[\protect\citeauthoryear{Codd}{Codd}{1972}]%
        {codd1972relational}
\bibfield{author}{\bibinfo{person}{E.~F. Codd}.}
  \bibinfo{year}{1972}\natexlab{}.
\newblock \bibinfo{booktitle}{\emph{Relational Completeness of Data Base
  Sublanguages}}.
\newblock \bibinfo{publisher}{IBM Corporation}.
\newblock


\bibitem[\protect\citeauthoryear{Cuoq, Monate, Pacalet, Prevosto, Regehr,
  Yakobowski, and Yang}{Cuoq et~al\mbox{.}}{2012}]%
        {Pascal2012}
\bibfield{author}{\bibinfo{person}{Pascal Cuoq}, \bibinfo{person}{Benjamin
  Monate}, \bibinfo{person}{Anne Pacalet}, \bibinfo{person}{Virgile Prevosto},
  \bibinfo{person}{John Regehr}, \bibinfo{person}{Boris Yakobowski}, {and}
  \bibinfo{person}{Xuejun Yang}.} \bibinfo{year}{2012}\natexlab{}.
\newblock \showarticletitle{Testing Static Analyzers with Randomly Generated
  Programs}. In \bibinfo{booktitle}{\emph{Proceedings of the 4th International
  Conference on NASA Formal Methods}} (Norfolk, VA)
  \emph{(\bibinfo{series}{NFM’12})}. \bibinfo{publisher}{Springer-Verlag},
  \bibinfo{address}{Berlin, Heidelberg}, \bibinfo{pages}{120–125}.
\newblock
\showISBNx{9783642288906}
\urldef\tempurl%
\url{https://doi.org/10.1007/978-3-642-28891-3_12}
\showDOI{\tempurl}


\bibitem[\protect\citeauthoryear{DB-Engines}{DB-Engines}{2019}]%
        {dbengines}
\bibfield{author}{\bibinfo{person}{DB-Engines}.}
  \bibinfo{year}{2019}\natexlab{}.
\newblock \bibinfo{title}{DB-Engines Ranking (July 2019)}.
\newblock
\newblock
\urldef\tempurl%
\url{https://db-engines.com/en/ranking}
\showURL{%
\tempurl}


\bibitem[\protect\citeauthoryear{Ding, Das, Wu, Chaudhuri, and Narasayya}{Ding
  et~al\mbox{.}}{2018}]%
        {Ding2018}
\bibfield{author}{\bibinfo{person}{Bailu Ding}, \bibinfo{person}{Sudipto Das},
  \bibinfo{person}{Wentao Wu}, \bibinfo{person}{Surajit Chaudhuri}, {and}
  \bibinfo{person}{Vivek Narasayya}.} \bibinfo{year}{2018}\natexlab{}.
\newblock \showarticletitle{Plan Stitch: Harnessing the Best of Many Plans}.
\newblock \bibinfo{journal}{\emph{Proc. VLDB Endow.}} \bibinfo{volume}{11},
  \bibinfo{number}{10} (\bibinfo{date}{June} \bibinfo{year}{2018}),
  \bibinfo{pages}{1123–1136}.
\newblock
\showISSN{2150-8097}
\urldef\tempurl%
\url{https://doi.org/10.14778/3231751.3231761}
\showDOI{\tempurl}


\bibitem[\protect\citeauthoryear{Elmasri and Navathe}{Elmasri and
  Navathe}{2017}]%
        {elmasri2017fundamentals}
\bibfield{author}{\bibinfo{person}{Ramez Elmasri} {and} \bibinfo{person}{Sham
  Navathe}.} \bibinfo{year}{2017}\natexlab{}.
\newblock \bibinfo{booktitle}{\emph{Fundamentals of database systems}}.
  Vol.~\bibinfo{volume}{7}.
\newblock \bibinfo{publisher}{Pearson}.
\newblock


\bibitem[\protect\citeauthoryear{Giakoumakis and Galindo-Legaria}{Giakoumakis
  and Galindo-Legaria}{2008}]%
        {giakoumakis2008testing}
\bibfield{author}{\bibinfo{person}{Leo Giakoumakis} {and}
  \bibinfo{person}{C{\'e}sar~A Galindo-Legaria}.}
  \bibinfo{year}{2008}\natexlab{}.
\newblock \showarticletitle{Testing SQL Server's Query Optimizer: Challenges,
  Techniques and Experiences.}
\newblock \bibinfo{journal}{\emph{IEEE Data Eng. Bull.}} \bibinfo{volume}{31},
  \bibinfo{number}{1} (\bibinfo{year}{2008}), \bibinfo{pages}{36--43}.
\newblock


\bibitem[\protect\citeauthoryear{Grabs, Herbert, and Zhang}{Grabs
  et~al\mbox{.}}{2008}]%
        {Torsten2008}
\bibfield{author}{\bibinfo{person}{Torsten Grabs}, \bibinfo{person}{Steve
  Herbert}, {and} \bibinfo{person}{Xin~(Shin) Zhang}.}
  \bibinfo{year}{2008}\natexlab{}.
\newblock \showarticletitle{Testing Challenges for Extending SQL Server’s
  Query Processor: A Case Study}. In \bibinfo{booktitle}{\emph{Proceedings of
  the 1st International Workshop on Testing Database Systems}} (Vancouver,
  British Columbia, Canada) \emph{(\bibinfo{series}{DBTest ’08})}.
  \bibinfo{publisher}{Association for Computing Machinery},
  \bibinfo{address}{New York, NY, USA}, Article \bibinfo{articleno}{Article 2},
  \bibinfo{numpages}{6}~pages.
\newblock
\showISBNx{9781605582337}
\urldef\tempurl%
\url{https://doi.org/10.1145/1385269.1385272}
\showDOI{\tempurl}


\bibitem[\protect\citeauthoryear{Graefe}{Graefe}{1993}]%
        {graefe1993query}
\bibfield{author}{\bibinfo{person}{Goetz Graefe}.}
  \bibinfo{year}{1993}\natexlab{}.
\newblock \showarticletitle{Query evaluation techniques for large databases}.
\newblock \bibinfo{journal}{\emph{ACM Computing Surveys (CSUR)}}
  \bibinfo{volume}{25}, \bibinfo{number}{2} (\bibinfo{year}{1993}),
  \bibinfo{pages}{73--169}.
\newblock


\bibitem[\protect\citeauthoryear{Gray, Sundaresan, Englert, Baclawski, and
  Weinberger}{Gray et~al\mbox{.}}{1994}]%
        {gray1994}
\bibfield{author}{\bibinfo{person}{Jim Gray}, \bibinfo{person}{Prakash
  Sundaresan}, \bibinfo{person}{Susanne Englert}, \bibinfo{person}{Ken
  Baclawski}, {and} \bibinfo{person}{Peter~J. Weinberger}.}
  \bibinfo{year}{1994}\natexlab{}.
\newblock \showarticletitle{Quickly Generating Billion-Record Synthetic
  Databases}.
\newblock \bibinfo{journal}{\emph{SIGMOD Rec.}} \bibinfo{volume}{23},
  \bibinfo{number}{2} (\bibinfo{date}{May} \bibinfo{year}{1994}),
  \bibinfo{pages}{243–252}.
\newblock
\showISSN{0163-5808}
\urldef\tempurl%
\url{https://doi.org/10.1145/191843.191886}
\showDOI{\tempurl}


\bibitem[\protect\citeauthoryear{Gu, Soliman, and Waas}{Gu
  et~al\mbox{.}}{2012}]%
        {Gu2012}
\bibfield{author}{\bibinfo{person}{Zhongxian Gu}, \bibinfo{person}{Mohamed~A.
  Soliman}, {and} \bibinfo{person}{Florian~M. Waas}.}
  \bibinfo{year}{2012}\natexlab{}.
\newblock \showarticletitle{Testing the Accuracy of Query Optimizers}. In
  \bibinfo{booktitle}{\emph{Proceedings of the Fifth International Workshop on
  Testing Database Systems}} (Scottsdale, Arizona)
  \emph{(\bibinfo{series}{DBTest '12})}. \bibinfo{publisher}{ACM},
  \bibinfo{address}{New York, NY, USA}, Article \bibinfo{articleno}{11},
  \bibinfo{numpages}{6}~pages.
\newblock
\showISBNx{978-1-4503-1429-9}
\urldef\tempurl%
\url{https://doi.org/10.1145/2304510.2304525}
\showDOI{\tempurl}


\bibitem[\protect\citeauthoryear{Guttman}{Guttman}{1984}]%
        {rtree}
\bibfield{author}{\bibinfo{person}{Antonin Guttman}.}
  \bibinfo{year}{1984}\natexlab{}.
\newblock \showarticletitle{R-Trees: A Dynamic Index Structure for Spatial
  Searching}. In \bibinfo{booktitle}{\emph{Proceedings of the 1984 ACM SIGMOD
  International Conference on Management of Data}} (Boston, Massachusetts)
  \emph{(\bibinfo{series}{SIGMOD ’84})}. \bibinfo{publisher}{Association for
  Computing Machinery}, \bibinfo{address}{New York, NY, USA},
  \bibinfo{pages}{47–57}.
\newblock
\showISBNx{0897911288}
\urldef\tempurl%
\url{https://doi.org/10.1145/602259.602266}
\showDOI{\tempurl}


\bibitem[\protect\citeauthoryear{Houkjær, Torp, and Wind}{Houkjær
  et~al\mbox{.}}{2006}]%
        {houkjaer2006}
\bibfield{author}{\bibinfo{person}{Kenneth Houkjær}, \bibinfo{person}{Kristian
  Torp}, {and} \bibinfo{person}{Rico Wind}.} \bibinfo{year}{2006}\natexlab{}.
\newblock \showarticletitle{Simple and Realistic Data Generation}. In
  \bibinfo{booktitle}{\emph{Proceedings of the 32nd International Conference on
  Very Large Data Bases}} (Seoul, Korea) \emph{(\bibinfo{series}{VLDB ’06})}.
  \bibinfo{publisher}{VLDB Endowment}, \bibinfo{pages}{1243–1246}.
\newblock


\bibitem[\protect\citeauthoryear{Howden}{Howden}{1978}]%
        {Howden:1978}
\bibfield{author}{\bibinfo{person}{William~E. Howden}.}
  \bibinfo{year}{1978}\natexlab{}.
\newblock \showarticletitle{Theoretical and Empirical Studies of Program
  Testing}. In \bibinfo{booktitle}{\emph{Proceedings of the 3rd International
  Conference on Software Engineering}} (Atlanta, Georgia, USA)
  \emph{(\bibinfo{series}{ICSE '78})}. \bibinfo{publisher}{IEEE Press},
  \bibinfo{address}{Piscataway, NJ, USA}, \bibinfo{pages}{305--311}.
\newblock


\bibitem[\protect\citeauthoryear{Jibson}{Jibson}{2016}]%
        {Jibson2016}
\bibfield{author}{\bibinfo{person}{Matt Jibson}.}
  \bibinfo{year}{2016}\natexlab{}.
\newblock \bibinfo{title}{Testing Random, Valid SQL in CockroachDB}.
\newblock
\newblock
\urldef\tempurl%
\url{https://www.cockroachlabs.com/blog/testing-random-valid-sql-in-cockroachdb/}
\showURL{%
\tempurl}


\bibitem[\protect\citeauthoryear{Jibson}{Jibson}{2019}]%
        {Jibson2019}
\bibfield{author}{\bibinfo{person}{Matt Jibson}.}
  \bibinfo{year}{2019}\natexlab{}.
\newblock \bibinfo{title}{SQLsmith: Randomized SQL Testing in CockroachDB}.
\newblock
\newblock
\urldef\tempurl%
\url{https://www.cockroachlabs.com/blog/sqlsmith-randomized-sql-testing/}
\showURL{%
\tempurl}


\bibitem[\protect\citeauthoryear{Jung, Hu, Arulraj, Kim, and Kang}{Jung
  et~al\mbox{.}}{2019}]%
        {apollo}
\bibfield{author}{\bibinfo{person}{Jinho Jung}, \bibinfo{person}{Hong Hu},
  \bibinfo{person}{Joy Arulraj}, \bibinfo{person}{Taesoo Kim}, {and}
  \bibinfo{person}{Woonhak Kang}.} \bibinfo{year}{2019}\natexlab{}.
\newblock \showarticletitle{APOLLO: Automatic Detection and Diagnosis of
  Performance Regressions in Database Systems}.
\newblock \bibinfo{journal}{\emph{Proc. VLDB Endow.}} \bibinfo{volume}{13},
  \bibinfo{number}{1} (\bibinfo{date}{Sept.} \bibinfo{year}{2019}),
  \bibinfo{pages}{57–70}.
\newblock
\showISSN{2150-8097}
\urldef\tempurl%
\url{https://doi.org/10.14778/3357377.3357382}
\showDOI{\tempurl}


\bibitem[\protect\citeauthoryear{Kapus and Cadar}{Kapus and Cadar}{2017}]%
        {Kapus2017}
\bibfield{author}{\bibinfo{person}{Timotej Kapus} {and}
  \bibinfo{person}{Cristian Cadar}.} \bibinfo{year}{2017}\natexlab{}.
\newblock \showarticletitle{Automatic Testing of Symbolic Execution Engines via
  Program Generation and Differential Testing}. In
  \bibinfo{booktitle}{\emph{Proceedings of the 32Nd IEEE/ACM International
  Conference on Automated Software Engineering}} (Urbana-Champaign, IL, USA)
  \emph{(\bibinfo{series}{ASE 2017})}. \bibinfo{publisher}{IEEE Press},
  \bibinfo{address}{Piscataway, NJ, USA}, \bibinfo{pages}{590--600}.
\newblock
\showISBNx{978-1-5386-2684-9}


\bibitem[\protect\citeauthoryear{Khalek, Elkarablieh, Laleye, and
  Khurshid}{Khalek et~al\mbox{.}}{2008}]%
        {khalek2008}
\bibfield{author}{\bibinfo{person}{S.~A. Khalek}, \bibinfo{person}{B.
  Elkarablieh}, \bibinfo{person}{Y.~O. Laleye}, {and} \bibinfo{person}{S.
  Khurshid}.} \bibinfo{year}{2008}\natexlab{}.
\newblock \showarticletitle{Query-Aware Test Generation Using a Relational
  Constraint Solver}. In \bibinfo{booktitle}{\emph{Proceedings of the 2008 23rd
  IEEE/ACM International Conference on Automated Software Engineering}}
  \emph{(\bibinfo{series}{ASE '08})}. \bibinfo{publisher}{IEEE Computer
  Society}, \bibinfo{address}{Washington, DC, USA}, \bibinfo{pages}{238--247}.
\newblock
\showISBNx{978-1-4244-2187-9}
\urldef\tempurl%
\url{https://doi.org/10.1109/ASE.2008.34}
\showDOI{\tempurl}


\bibitem[\protect\citeauthoryear{Le, Afshari, and Su}{Le et~al\mbox{.}}{2014}]%
        {Le2014}
\bibfield{author}{\bibinfo{person}{Vu Le}, \bibinfo{person}{Mehrdad Afshari},
  {and} \bibinfo{person}{Zhendong Su}.} \bibinfo{year}{2014}\natexlab{}.
\newblock \showarticletitle{Compiler Validation via Equivalence Modulo Inputs}.
  In \bibinfo{booktitle}{\emph{Proceedings of the 35th ACM SIGPLAN Conference
  on Programming Language Design and Implementation}} (Edinburgh, United
  Kingdom) \emph{(\bibinfo{series}{PLDI '14})}. \bibinfo{publisher}{ACM},
  \bibinfo{address}{New York, NY, USA}, \bibinfo{pages}{216--226}.
\newblock
\showISBNx{978-1-4503-2784-8}
\urldef\tempurl%
\url{https://doi.org/10.1145/2594291.2594334}
\showDOI{\tempurl}


\bibitem[\protect\citeauthoryear{Lo, Binnig, Kossmann, Tamer~{\"O}zsu, and
  Hon}{Lo et~al\mbox{.}}{2010}]%
        {Lo2010framework}
\bibfield{author}{\bibinfo{person}{Eric Lo}, \bibinfo{person}{Carsten Binnig},
  \bibinfo{person}{Donald Kossmann}, \bibinfo{person}{M. Tamer~{\"O}zsu}, {and}
  \bibinfo{person}{Wing-Kai Hon}.} \bibinfo{year}{2010}\natexlab{}.
\newblock \showarticletitle{A framework for testing DBMS features}.
\newblock \bibinfo{journal}{\emph{The VLDB Journal}} \bibinfo{volume}{19},
  \bibinfo{number}{2} (\bibinfo{date}{01 Apr} \bibinfo{year}{2010}),
  \bibinfo{pages}{203--230}.
\newblock
\showISSN{0949-877X}
\urldef\tempurl%
\url{https://doi.org/10.1007/s00778-009-0157-y}
\showDOI{\tempurl}


\bibitem[\protect\citeauthoryear{Marcus, Negi, Mao, Zhang, Alizadeh, Kraska,
  Papaemmanouil, and Tatbul}{Marcus et~al\mbox{.}}{2019}]%
        {Marcus2019}
\bibfield{author}{\bibinfo{person}{Ryan Marcus}, \bibinfo{person}{Parimarjan
  Negi}, \bibinfo{person}{Hongzi Mao}, \bibinfo{person}{Chi Zhang},
  \bibinfo{person}{Mohammad Alizadeh}, \bibinfo{person}{Tim Kraska},
  \bibinfo{person}{Olga Papaemmanouil}, {and} \bibinfo{person}{Nesime Tatbul}.}
  \bibinfo{year}{2019}\natexlab{}.
\newblock \showarticletitle{Neo: A Learned Query Optimizer}.
\newblock \bibinfo{journal}{\emph{Proc. VLDB Endow.}} \bibinfo{volume}{12},
  \bibinfo{number}{11} (\bibinfo{date}{July} \bibinfo{year}{2019}),
  \bibinfo{pages}{1705–1718}.
\newblock
\showISSN{2150-8097}
\urldef\tempurl%
\url{https://doi.org/10.14778/3342263.3342644}
\showDOI{\tempurl}


\bibitem[\protect\citeauthoryear{McKeeman}{McKeeman}{1998}]%
        {mckeeman1998differential}
\bibfield{author}{\bibinfo{person}{William~M McKeeman}.}
  \bibinfo{year}{1998}\natexlab{}.
\newblock \showarticletitle{Differential testing for software}.
\newblock \bibinfo{journal}{\emph{Digital Technical Journal}}
  \bibinfo{volume}{10}, \bibinfo{number}{1} (\bibinfo{year}{1998}),
  \bibinfo{pages}{100--107}.
\newblock


\bibitem[\protect\citeauthoryear{Mishra, Koudas, and Zuzarte}{Mishra
  et~al\mbox{.}}{2008}]%
        {Mishra2008}
\bibfield{author}{\bibinfo{person}{Chaitanya Mishra}, \bibinfo{person}{Nick
  Koudas}, {and} \bibinfo{person}{Calisto Zuzarte}.}
  \bibinfo{year}{2008}\natexlab{}.
\newblock \showarticletitle{Generating Targeted Queries for Database Testing}.
  In \bibinfo{booktitle}{\emph{Proceedings of the 2008 ACM SIGMOD International
  Conference on Management of Data}} (Vancouver, Canada)
  \emph{(\bibinfo{series}{SIGMOD '08})}. \bibinfo{publisher}{ACM},
  \bibinfo{address}{New York, NY, USA}, \bibinfo{pages}{499--510}.
\newblock
\showISBNx{978-1-60558-102-6}
\urldef\tempurl%
\url{https://doi.org/10.1145/1376616.1376668}
\showDOI{\tempurl}


\bibitem[\protect\citeauthoryear{Neufeld, Moerkotte, and Lockemann}{Neufeld
  et~al\mbox{.}}{1993}]%
        {neufeld1993}
\bibfield{author}{\bibinfo{person}{Andrea Neufeld}, \bibinfo{person}{Guido
  Moerkotte}, {and} \bibinfo{person}{Peter~C. Lockemann}.}
  \bibinfo{year}{1993}\natexlab{}.
\newblock \showarticletitle{Generating Consistent Test Data: Restricting the
  Search Space by a Generator Formula}.
\newblock \bibinfo{journal}{\emph{The VLDB Journal}} \bibinfo{volume}{2},
  \bibinfo{number}{2} (\bibinfo{date}{April} \bibinfo{year}{1993}),
  \bibinfo{pages}{173–214}.
\newblock
\showISSN{1066-8888}


\bibitem[\protect\citeauthoryear{Neumann and Radke}{Neumann and Radke}{2018}]%
        {Neumann2018}
\bibfield{author}{\bibinfo{person}{Thomas Neumann} {and}
  \bibinfo{person}{Bernhard Radke}.} \bibinfo{year}{2018}\natexlab{}.
\newblock \showarticletitle{Adaptive Optimization of Very Large Join Queries}.
  In \bibinfo{booktitle}{\emph{Proceedings of the 2018 International Conference
  on Management of Data}} (Houston, TX, USA) \emph{(\bibinfo{series}{SIGMOD
  ’18})}. \bibinfo{publisher}{Association for Computing Machinery},
  \bibinfo{address}{New York, NY, USA}, \bibinfo{pages}{677–692}.
\newblock
\showISBNx{9781450347037}
\urldef\tempurl%
\url{https://doi.org/10.1145/3183713.3183733}
\showDOI{\tempurl}


\bibitem[\protect\citeauthoryear{Overflow}{Overflow}{2019}]%
        {stackoverflowsurvey}
\bibfield{author}{\bibinfo{person}{Stack Overflow}.}
  \bibinfo{year}{2019}\natexlab{}.
\newblock \bibinfo{title}{Developer Survey Results 2019}.
\newblock
\newblock
\urldef\tempurl%
\url{https://insights.stackoverflow.com/survey/2019}
\showURL{%
\tempurl}


\bibitem[\protect\citeauthoryear{Pavlo and Aslett}{Pavlo and Aslett}{2016}]%
        {newsql}
\bibfield{author}{\bibinfo{person}{Andrew Pavlo} {and} \bibinfo{person}{Matthew
  Aslett}.} \bibinfo{year}{2016}\natexlab{}.
\newblock \showarticletitle{What's Really New with NewSQL?}
\newblock \bibinfo{journal}{\emph{SIGMOD Rec.}} \bibinfo{volume}{45},
  \bibinfo{number}{2} (\bibinfo{date}{Sept.} \bibinfo{year}{2016}),
  \bibinfo{pages}{45–55}.
\newblock
\showISSN{0163-5808}
\urldef\tempurl%
\url{https://doi.org/10.1145/3003665.3003674}
\showDOI{\tempurl}


\bibitem[\protect\citeauthoryear{Poess and Stephens}{Poess and
  Stephens}{2004}]%
        {qgen}
\bibfield{author}{\bibinfo{person}{Meikel Poess} {and} \bibinfo{person}{John~M.
  Stephens}.} \bibinfo{year}{2004}\natexlab{}.
\newblock \showarticletitle{Generating Thousand Benchmark Queries in Seconds}.
  In \bibinfo{booktitle}{\emph{Proceedings of the Thirtieth International
  Conference on Very Large Data Bases - Volume 30}} (Toronto, Canada)
  \emph{(\bibinfo{series}{VLDB ’04})}. \bibinfo{publisher}{VLDB Endowment},
  \bibinfo{pages}{1045–1053}.
\newblock
\showISBNx{0120884690}


\bibitem[\protect\citeauthoryear{Regehr, Chen, Cuoq, Eide, Ellison, and
  Yang}{Regehr et~al\mbox{.}}{2012}]%
        {creduce}
\bibfield{author}{\bibinfo{person}{John Regehr}, \bibinfo{person}{Yang Chen},
  \bibinfo{person}{Pascal Cuoq}, \bibinfo{person}{Eric Eide},
  \bibinfo{person}{Chucky Ellison}, {and} \bibinfo{person}{Xuejun Yang}.}
  \bibinfo{year}{2012}\natexlab{}.
\newblock \showarticletitle{Test-Case Reduction for C Compiler Bugs}. In
  \bibinfo{booktitle}{\emph{Proceedings of the 33rd ACM SIGPLAN Conference on
  Programming Language Design and Implementation}} (Beijing, China)
  \emph{(\bibinfo{series}{PLDI ’12})}. \bibinfo{publisher}{Association for
  Computing Machinery}, \bibinfo{address}{New York, NY, USA},
  \bibinfo{pages}{335–346}.
\newblock
\showISBNx{9781450312059}
\urldef\tempurl%
\url{https://doi.org/10.1145/2254064.2254104}
\showDOI{\tempurl}


\bibitem[\protect\citeauthoryear{Rigger}{Rigger}{2019}]%
        {notnullnotnullbug}
\bibfield{author}{\bibinfo{person}{Manuel Rigger}.}
  \bibinfo{year}{2019}\natexlab{}.
\newblock \bibinfo{title}{LEFT JOIN in view malfunctions with NOTNULL}.
\newblock
\newblock
\urldef\tempurl%
\url{https://www.sqlite.org/src/tktview?name=c31034044b}
\showURL{%
\tempurl}


\bibitem[\protect\citeauthoryear{Rigger and Su}{Rigger and Su}{2020a}]%
        {norecartifact}
\bibfield{author}{\bibinfo{person}{Manuel Rigger} {and}
  \bibinfo{person}{Zhendong Su}.} \bibinfo{year}{2020}\natexlab{a}.
\newblock \bibinfo{title}{{ESEC/FSE 20 Artifact for "Detecting Optimization
  Bugs in Database Engines via Non-Optimizing Reference Engine Construction"}}.
\newblock
\newblock
\urldef\tempurl%
\url{https://doi.org/10.5281/zenodo.3947858}
\showDOI{\tempurl}


\bibitem[\protect\citeauthoryear{Rigger and Su}{Rigger and Su}{2020b}]%
        {pqs}
\bibfield{author}{\bibinfo{person}{Manuel Rigger} {and}
  \bibinfo{person}{Zhendong Su}.} \bibinfo{year}{2020}\natexlab{b}.
\newblock \bibinfo{title}{{Testing Database Engines via Pivoted Query
  Synthesis}}.
\newblock
\newblock


\bibitem[\protect\citeauthoryear{Segura and Zhou}{Segura and Zhou}{2018}]%
        {Sergio2018}
\bibfield{author}{\bibinfo{person}{Sergio Segura} {and}
  \bibinfo{person}{Zhi~Quan Zhou}.} \bibinfo{year}{2018}\natexlab{}.
\newblock \showarticletitle{Metamorphic Testing 20 Years Later: A Hands-on
  Introduction}. In \bibinfo{booktitle}{\emph{Proceedings of the 40th
  International Conference on Software Engineering: Companion Proceeedings}}
  (Gothenburg, Sweden) \emph{(\bibinfo{series}{ICSE ’18})}.
  \bibinfo{publisher}{Association for Computing Machinery},
  \bibinfo{address}{New York, NY, USA}, \bibinfo{pages}{538–539}.
\newblock
\showISBNx{9781450356633}
\urldef\tempurl%
\url{https://doi.org/10.1145/3183440.3183468}
\showDOI{\tempurl}


\bibitem[\protect\citeauthoryear{Selinger, Astrahan, Chamberlin, Lorie, and
  Price}{Selinger et~al\mbox{.}}{1979}]%
        {selinger1979}
\bibfield{author}{\bibinfo{person}{P.~Griffiths Selinger},
  \bibinfo{person}{M.~M. Astrahan}, \bibinfo{person}{D.~D. Chamberlin},
  \bibinfo{person}{R.~A. Lorie}, {and} \bibinfo{person}{T.~G. Price}.}
  \bibinfo{year}{1979}\natexlab{}.
\newblock \showarticletitle{Access Path Selection in a Relational Database
  Management System}. In \bibinfo{booktitle}{\emph{Proceedings of the 1979 ACM
  SIGMOD International Conference on Management of Data}} (Boston,
  Massachusetts) \emph{(\bibinfo{series}{SIGMOD ’79})}.
  \bibinfo{publisher}{Association for Computing Machinery},
  \bibinfo{address}{New York, NY, USA}, \bibinfo{pages}{23–34}.
\newblock
\showISBNx{089791001X}
\urldef\tempurl%
\url{https://doi.org/10.1145/582095.582099}
\showDOI{\tempurl}


\bibitem[\protect\citeauthoryear{Seltenreich}{Seltenreich}{2019}]%
        {sqlsmith}
\bibfield{author}{\bibinfo{person}{Andreas Seltenreich}.}
  \bibinfo{year}{2019}\natexlab{}.
\newblock \bibinfo{title}{SQLSmith}.
\newblock
\newblock
\urldef\tempurl%
\url{https://github.com/anse1/sqlsmith}
\showURL{%
\tempurl}


\bibitem[\protect\citeauthoryear{Slutz}{Slutz}{1998}]%
        {slutz1998massive}
\bibfield{author}{\bibinfo{person}{Donald~R Slutz}.}
  \bibinfo{year}{1998}\natexlab{}.
\newblock \showarticletitle{Massive stochastic testing of SQL}. In
  \bibinfo{booktitle}{\emph{VLDB}}, Vol.~\bibinfo{volume}{98}.
  \bibinfo{pages}{618--622}.
\newblock


\bibitem[\protect\citeauthoryear{SQLite3}{SQLite3}{2020a}]%
        {sqlitegeneratedcolumns}
\bibfield{author}{\bibinfo{person}{SQLite3}.} \bibinfo{year}{2020}\natexlab{a}.
\newblock \bibinfo{title}{Generated Columns}.
\newblock
\newblock
\urldef\tempurl%
\url{https://sqlite.org/gencol.html}
\showURL{%
\tempurl}


\bibitem[\protect\citeauthoryear{SQLite3}{SQLite3}{2020b}]%
        {sqlitetesting}
\bibfield{author}{\bibinfo{person}{SQLite3}.} \bibinfo{year}{2020}\natexlab{b}.
\newblock \bibinfo{title}{How SQLite Is Tested}.
\newblock
\newblock
\urldef\tempurl%
\url{https://www.sqlite.org/testing.html}
\showURL{%
\tempurl}


\bibitem[\protect\citeauthoryear{SQLite3}{SQLite3}{2020c}]%
        {sqlitemostdeployed}
\bibfield{author}{\bibinfo{person}{SQLite3}.} \bibinfo{year}{2020}\natexlab{c}.
\newblock \bibinfo{title}{Most Widely Deployed and Used Database Engine}.
\newblock
\newblock
\urldef\tempurl%
\url{https://www.sqlite.org/mostdeployed.html}
\showURL{%
\tempurl}


\bibitem[\protect\citeauthoryear{SQLite3}{SQLite3}{2020d}]%
        {sqliteoptimizer}
\bibfield{author}{\bibinfo{person}{SQLite3}.} \bibinfo{year}{2020}\natexlab{d}.
\newblock \bibinfo{title}{The SQLite Query Optimizer Overview}.
\newblock
\newblock
\urldef\tempurl%
\url{https://www.sqlite.org/optoverview.html}
\showURL{%
\tempurl}


\bibitem[\protect\citeauthoryear{SQLite3}{SQLite3}{2020e}]%
        {assertsqlite}
\bibfield{author}{\bibinfo{person}{SQLite3}.} \bibinfo{year}{2020}\natexlab{e}.
\newblock \bibinfo{title}{The Use Of assert() In SQLite}.
\newblock
\newblock
\urldef\tempurl%
\url{https://www.sqlite.org/assert.html}
\showURL{%
\tempurl}


\bibitem[\protect\citeauthoryear{Taft, Sharif, Matei, VanBenschoten, Lewis,
  Grieger, Niemi, Woods, Birzin, Poss, Bardea, Ranade, Darnell, Gruneir,
  Jaffray, Zhang, and Mattis}{Taft et~al\mbox{.}}{2020}]%
        {cockroachdb}
\bibfield{author}{\bibinfo{person}{Rebecca Taft}, \bibinfo{person}{Irfan
  Sharif}, \bibinfo{person}{Andrei Matei}, \bibinfo{person}{Nathan
  VanBenschoten}, \bibinfo{person}{Jordan Lewis}, \bibinfo{person}{Tobias
  Grieger}, \bibinfo{person}{Kai Niemi}, \bibinfo{person}{Andy Woods},
  \bibinfo{person}{Anne Birzin}, \bibinfo{person}{Raphael Poss},
  \bibinfo{person}{Paul Bardea}, \bibinfo{person}{Amruta Ranade},
  \bibinfo{person}{Ben Darnell}, \bibinfo{person}{Bram Gruneir},
  \bibinfo{person}{Justin Jaffray}, \bibinfo{person}{Lucy Zhang}, {and}
  \bibinfo{person}{Peter Mattis}.} \bibinfo{year}{2020}\natexlab{}.
\newblock \showarticletitle{CockroachDB: The Resilient Geo-Distributed SQL
  Database}. In \bibinfo{booktitle}{\emph{Proceedings of the 2020 ACM SIGMOD
  International Conference on Management of Data}} (Portland, OR, USA)
  \emph{(\bibinfo{series}{SIGMOD ’20})}. \bibinfo{publisher}{International
  Foundation for Autonomous Agents and Multiagent Systems},
  \bibinfo{address}{Richland, SC}, \bibinfo{pages}{1493–1509}.
\newblock
\showISBNx{9781450367356}
\urldef\tempurl%
\url{https://doi.org/10.1145/3318464.3386134}
\showDOI{\tempurl}


\bibitem[\protect\citeauthoryear{{Tencent Blade Team}}{{Tencent Blade
  Team}}{2019}]%
        {magellan}
\bibfield{author}{\bibinfo{person}{{Tencent Blade Team}}.}
  \bibinfo{year}{2019}\natexlab{}.
\newblock \bibinfo{title}{Magellan 2.0}.
\newblock
\newblock
\urldef\tempurl%
\url{https://blade.tencent.com/magellan2/index_en.html}
\showURL{%
\tempurl}


\bibitem[\protect\citeauthoryear{Vartak, Raghavan, and Rundensteiner}{Vartak
  et~al\mbox{.}}{2010}]%
        {qrelx}
\bibfield{author}{\bibinfo{person}{Manasi Vartak}, \bibinfo{person}{Venkatesh
  Raghavan}, {and} \bibinfo{person}{Elke~A. Rundensteiner}.}
  \bibinfo{year}{2010}\natexlab{}.
\newblock \showarticletitle{QRelX: Generating Meaningful Queries That Provide
  Cardinality Assurance}. In \bibinfo{booktitle}{\emph{Proceedings of the 2010
  ACM SIGMOD International Conference on Management of Data}} (Indianapolis,
  Indiana, USA) \emph{(\bibinfo{series}{SIGMOD ’10})}.
  \bibinfo{publisher}{Association for Computing Machinery},
  \bibinfo{address}{New York, NY, USA}, \bibinfo{pages}{1215–1218}.
\newblock
\showISBNx{9781450300322}
\urldef\tempurl%
\url{https://doi.org/10.1145/1807167.1807323}
\showDOI{\tempurl}


\bibitem[\protect\citeauthoryear{Winslett and Braganholo}{Winslett and
  Braganholo}{2019}]%
        {richardhippinterview}
\bibfield{author}{\bibinfo{person}{Marianne Winslett} {and}
  \bibinfo{person}{Vanessa Braganholo}.} \bibinfo{year}{2019}\natexlab{}.
\newblock \showarticletitle{Richard Hipp Speaks Out on SQLite}.
\newblock \bibinfo{journal}{\emph{SIGMOD Rec.}} \bibinfo{volume}{48},
  \bibinfo{number}{2} (\bibinfo{date}{Dec.} \bibinfo{year}{2019}),
  \bibinfo{pages}{39–46}.
\newblock
\showISSN{0163-5808}
\urldef\tempurl%
\url{https://doi.org/10.1145/3377330.3377338}
\showDOI{\tempurl}


\bibitem[\protect\citeauthoryear{Wu, Jindal, Amizadeh, Patel, Le, Qiao, and
  Rao}{Wu et~al\mbox{.}}{2018}]%
        {Wu2018}
\bibfield{author}{\bibinfo{person}{Chenggang Wu}, \bibinfo{person}{Alekh
  Jindal}, \bibinfo{person}{Saeed Amizadeh}, \bibinfo{person}{Hiren Patel},
  \bibinfo{person}{Wangchao Le}, \bibinfo{person}{Shi Qiao}, {and}
  \bibinfo{person}{Sriram Rao}.} \bibinfo{year}{2018}\natexlab{}.
\newblock \showarticletitle{Towards a Learning Optimizer for Shared Clouds}.
\newblock \bibinfo{journal}{\emph{Proc. VLDB Endow.}} \bibinfo{volume}{12},
  \bibinfo{number}{3} (\bibinfo{date}{Nov.} \bibinfo{year}{2018}),
  \bibinfo{pages}{210–222}.
\newblock
\showISSN{2150-8097}
\urldef\tempurl%
\url{https://doi.org/10.14778/3291264.3291267}
\showDOI{\tempurl}


\bibitem[\protect\citeauthoryear{Yang, Chen, Eide, and Regehr}{Yang
  et~al\mbox{.}}{2011}]%
        {Yang2011}
\bibfield{author}{\bibinfo{person}{Xuejun Yang}, \bibinfo{person}{Yang Chen},
  \bibinfo{person}{Eric Eide}, {and} \bibinfo{person}{John Regehr}.}
  \bibinfo{year}{2011}\natexlab{}.
\newblock \showarticletitle{Finding and Understanding Bugs in C Compilers}. In
  \bibinfo{booktitle}{\emph{Proceedings of the 32Nd ACM SIGPLAN Conference on
  Programming Language Design and Implementation}} (San Jose, California, USA)
  \emph{(\bibinfo{series}{PLDI '11})}. \bibinfo{publisher}{ACM},
  \bibinfo{address}{New York, NY, USA}, \bibinfo{pages}{283--294}.
\newblock
\showISBNx{978-1-4503-0663-8}
\urldef\tempurl%
\url{https://doi.org/10.1145/1993498.1993532}
\showDOI{\tempurl}


\end{thebibliography}

\end{document}